\documentclass{aa}
\usepackage{amsmath}

\usepackage[svgnames]{xcolor}
\definecolor{blue}{RGB}{0,0,255}
\usepackage{caption}
\usepackage[font={color=blue},figurename=Fig.,labelfont={it}]{caption}
\usepackage{bm}
\usepackage{subcaption}
\usepackage[varg]{txfonts}
\usepackage{graphicx}
\usepackage{natbib}
\usepackage{pdfpages}
\usepackage{wasysym}
\usepackage{rotating}
\bibpunct{(}{)}{;}{a}{}{,} % to follow the A&A style
\usepackage{hyperref}
\usepackage{indentfirst}
\usepackage{float}
\usepackage{lipsum} 

\hypersetup{
    colorlinks=true,                          
    linkcolor=blue, 
    citecolor=blue, 
    urlcolor=blue,
    linktoc=all
}

\newcommand{\Figure}{{\color{blue}{Figure\ }}}
\newcommand{\Figures}{{\color{blue}{Figures\ }}}

\begin{document}
\title{On the Nature and Origin of Atmospheric Annual and Semi-annual Oscillations}
\author{V. Courtillot\inst{1}
\and J-L. Le Mouël \inst{1}
\and F. Lopes\inst{1}
\and D. Gibert\inst{2}
}
\institute{{Universit\'e Paris Cité, Institut de Physique du globe de Paris, CNRS UMR 7154, F-75005 Paris, France}
\and {LGL-TPE - Laboratoire de Géologie de Lyon - Terre, Planètes, Environnement, Lyon, France}}

\date{}

\abstract {This paper proposes a joint analysis of variations of global sea-level pressure (\textbf{SLP}) and of Earth’s rotation \textbf{RP}, expressed as the coordinates of the rotation pole ($m_1$, $m_2$) and length of day (\textbf{lod}). We retain iterative singular spectrum analysis (\textbf{iSSA}) as the main tool to extract the trend, periods and quasi periods in the data time series. \textbf{SLP }components are a weak trend, eleven quasi-periodic or periodic components ($\sim$130, 90, 50, 22, 15, 4, 1.8 yr), an annual cycle and its first three harmonics. These periods are characteristic of the space-time evolution of the Earth’s rotation axis (\textbf{RP}) and are present in many characteristic features of solar and terrestrial physics. The amplitudes of the annual \textbf{SLP }component and its 3 first harmonics decrease from 93 hPa for the annual to 21 hPa for the third harmonic. In contrast, the components with pseudo-periods longer than a year range between 0.2 and 0.5 hPa. The trend (21 hPa) could be part of the 90 yr Gleissberg cycle. We focus mainly on the annual and to a lesser extent semi-annual components. The annual \textbf{RP}  and \textbf{SLP} components have a phase lag of 152 days (half he Euler period) and they have similar envelopes, offset by about 40 years: polar motion leads atmospheric pressure. Maps of the first three components of \textbf{SLP} (that together comprise more than 85\% of the data variance) reveal interesting symmetries. The trend is very stable and forms a triskel structure that can be modeled as Taylor-Couette flow of mode 3. The annual component is characterized by a large negative anomaly extending over Eurasia in the \textbf{NH} Summer (and the opposite in the \textbf{NH} Winter) and three large positive anomalies over Australia and the southern tips of South America and South Africa in the \textbf{SH} Spring (and the opposite in the \textbf{SH} Autumn) forming a triskel. The semi-annual component is characterized by three positive anomalies (an irregular triskel) in the \textbf{NH} Spring and Autumn (and the opposite in the \textbf{NH} Summer and Winter), and in the \textbf{SH} Spring and Autumn by a strong stable pattern consisting of three large negative anomalies forming a clear triskel within the 40$^{\circ}$-60$^{\circ}$ annulus formed by the southern oceans. A large positive anomaly centered over Antarctica, with its maximum displaced toward Australia, and a smaller one centered over Southern Africa complement the pattern. Analysis of \textbf{SSA} components of global sea level pressure shows a rather simple spatial distribution with the principal forcing factor being changes in parameters of the Earth’s rotation pole and velocity (measured through length of day). The flow can probably best be modeled as a set of coaxial cylinders arranged in groups of three (triskels) or four and controlled by Earth topography and continent/ocean geography. Flow patterns suggested by maps of the three main \textbf{SSA} components of \textbf{SLP} (trend, annual and semi-annual) are suggestive of Taylor-Couette flow.  The envelopes of the annual components of \textbf{SLP} and \textbf{RP} are offset by four decades and there are indications that causality is present in the sense from changes in Earth rotation axis force pressure variations. }

\titlerunning{SLP Annual and Semi-annual Oscillations }
\maketitle

\section{Introduction} 
    This paper is an attempt to obtain better constraints on the forcings of the trend and annual components of both global sea-level pressure and variations in the Earth's rotation, and to test the hypothesis that there might be a causal link between them.\\
    
    \cite{Lopes2022a} undertook the singular spectral analysis (\textbf{SSA}) of the evolution of mean sea-level atmospheric pressure (\textbf{SLP}) since 1850. In addition to dominant trends, a eleven quasi-periodic components were identified, with (pseudo-) periods of $\sim$130, 90, 50, 22, 15, 4, 1.8, 1, 0.5, 0.33, and 0.25 years, corresponding to the Jose, Gleissberg, Hale and Schwabe cycles, to the annual cycle and its first three harmonics. These periods are already known to be characteristic of the space-time evolution of the Earth's rotation axis: the rotation pole (\textbf{RP}) also undergoes periodic motions longer than 1 year, at least up to the Gleissberg $\sim$90 yr cycle  \citep{Chandler1891a,Chandler1891b,Markowitz1968,Kirov2002,Lambeck2005,Zotov2012,Chao2014,Zotov2016,Lopes2017,LeMouel2021,Lopes2021,Lopes2022a}. They are encountered in solar physics \citep{Gleissberg1939, Jose1965, Coles1980, Charvatova1991, Scafetta2010,LeMouel2017, Usoskin2017a, Scafetta2020a, Courtillot2021, Scafetta2021} and terrestrial climate \citep{Wood1974, Morth1979, Morner1984, Schlesinger1994, Lau1995, Courtillot2007, Courtillot2013, LeMouel2019a, Scafetta2020b, Connolly2021}. \\
    
    The rotation velocity is usually expressed as the length of day (\textbf{lod}); it contains periods of 1 year and shorter, but also (though they are weaker) longer periods from 11 yr to 18.6 yr  \citep{Guinot1973, Ray2014, LeMouel2019b, Dumont2021, Petrosino2022}. \cite{Lopes2022a} showed that the secular variation of \textbf{lod} since 1846 \cite{Stephenson1984} and \textbf{RP} have been carried by an oscillation whose period fits the Gleissberg cycle. \\
    
   \textbf{ RP} consists to first order of 3 \textbf{SSA} components that together amount to more than 85\%  of the total signal variance \cite{Lopes2017}: the \cite{Markowitz1968} drift, the \cite{Chandler1891a,Chandler1891b} free oscillation (actually a double component with periods  $\sim$433 and  $\sim$434 days), and the forced annual oscillation. Polar drift does not have a universally accepted explanation \citep{Stoyko1968, Hulot1996, Markowitz2013, Deng2021}. The two other oscillatory components (Chandler and annual) are obtained using the Liouville-Euler system, that describes the motion of a spherical rotating solid body (\textit{e.g.} \cite{Lambeck2005}, chapter 3, system 3.2.9). \\
   
   Polar drift has been discussed in \cite{LeMouel2021} and \cite{Lopes2022a} and the free Chandler wobble in \cite{Lopes2021}. In the present paper, we focus on the third major component, the forced annual oscillation. This oscillation is often assumed to be forced by the Earth's fluid envelopes, hence also by climate variations (\textit{e.g.} \cite{Gross2003}, \cite{Lambeck2005} chapter 7, \cite{Bizouard2010, Chen2019}).  The fact that the observed Chandler period (433-434 days) is not equal to the theoretical value of the Euler period (306 days) has been interpreted in terms of Earth elasticity. The annual period is forced by interactions with fluids. Variations in the Sun-Earth distance, hence of the corresponding gravitational forces, displace the fluid atmosphere and ocean; exchanges in angular moment affect \textbf{RP} and \textbf{lod}. The annual oscillation of the rotation pole \textbf{RP} (coordinates $m_1$ and $m_2$) would therefore be due to the presence of the fluid envelopes. If Earth were devoid of fluid envelopes, the annual oscillation should not exist, which would contradict both the theory and observations of motion of a Lagrange top \citep{Landau1984}. \cite{Wilson1976} recall that \cite{Spitaler1897, Spitaler1901} “{\color{blue}\textit{demonstrated that the annual wobble was forced, at least in part, by the seasonal migration of air masses on and off the Asian continent}}”. \cite{Jeffreys1916} showed that the annual fluctuation in water storage on the continents was also important. \cite{Munk1961} re-examined the sources of annual wobble excitation and concluded that the air mass effect accounted for much but not all of the annual wobble, and water storage did not explain the remainder. Reservations appear in chapter 7 of \cite{Lambeck2005}, Seasonal Variations, page 146, who writes: “{\color{blue}{\textit{The principal seasonal oscillation in the wobble is the annual term which has generally been attributed to a geographical redistribution of mass associated with meteorological causes. Jeffreys, in 1916, first attempted a detailed quantitative evaluation of this excitation function by considering the contributions from atmospheric and oceanic motion, of precipitation, of vegetation and of polar ice. Jeffreys concluded that these factors explain the observed annual polar motion, a conclusion still valid today, although the quantitative comparisons between the observed and computed annual components of the pole path are still not satisfactory}}}”. In summary, the interpretation of the forced annual component of polar motion is still hypothetical and fails to be validated by a numerical model. \\
   
   In a previous analysis of global sea-level pressure (\textbf{SLP}), we found that trends since 1950 were very stable in time and space \citep{Lopes2022b} and were organized in a dominant 3-fold symmetry about the rotation axis in the northern hemisphere (\textbf{NH}) and a 3 or 4-fold symmetry in the southern hemisphere (\textbf{SH}) \citep{Lopes2022b}. These features could be interpreted as resulting from Taylor-Couette flow. In this paper, we return to the pressure data and focus on the annual oscillation.
    
\section{The Pressure Data and Method of Analysis}
    The pressure data are maintained by the Met Office Hadley Centre\footnote{\scriptsize https://www.metoffice.gov.uk/hadobs/hadslp2/data/download.html} and can be accessed as maps of global pressure, every month from 1850 up to the Present. The sampling is 5$^{\circ}$ x 5$^{\circ}$ ; the data we use are labeled \textbf{HadSLP2r}. \cite{Allan2006} have built this series starting with ground observations whose number and location are shown in their Figure 2. These are subjected to a quality check, corrected for various local effects then homogenized with the Empirical Mode Decomposition (\textbf{EMD}) filter. \\
    
    As in \cite{Lopes2022b}, we have used the iterative Singular Spectrum Analysis algorithm (\textbf{iSSA}) to extract the components, mainly the annual one, in each grid cell and as a function of time. For the \textbf{iSSA} method, see \cite{Golyandina2013}, for properties of the Hankel and Toeplitz matrices, see \cite{Lemmerling2001} and for the Singular Value Decomposition (\textbf{SVD}) algorithm, see \cite{Golub1971}. For summaries of how we use them, the reader is referred to \cite{Lopes2021}.
    
\section{The SSA Pressure Components}
    We obtain the following components: \\
    
    The trend (from $\sim$ 1009.15 hPa in 1850 to $\sim$ 1008.35 hPa at present) is the first and largest \textbf{SSA} component. It represents more than 70\% of the total variance of the original series. The sequence of the quasi-periodic components is, in decreasing order of periods (amplitudes are in hectoPascal, hPa) :    

\begin{itemize}
    \item  $\sim$ 130 years ($\sim$0.7 hPa). We recognize the \cite{Jose1965} cycle,
    \item  $\sim$ 90 yr ($\sim$21 hPa). We recognize the \cite{Gleissberg1939} cycle \citep{LeMouel2017},
     \item  $\sim$ 50 yr ($\sim$0.2 hPa),
     \item  $\sim$ 22 yr ($\sim$0.50 hPa). We recognize the Hale cycle \citep{Usoskin2017b},
     \item  $\sim$ 15 yr ($\sim$0.2 hPa). We may recognize an upper bound of the \cite{Schwabe1844}, cycle,
    \item  $\sim$ 4 yr ($\sim$0.3 hPa),
    \item  $\sim$ 1.8 yr ($\sim$0.3 hPa).
\end{itemize}

Then: 
\begin{itemize}
    \item  1 yr ( $\sim$93 hPa).
     \item  0.5 yr ($\sim$65 hPa).
     \item  0.33 yr ($\sim$44 hPa).
     \item  0.25 yr ($\sim$21 hPa).
\end{itemize}

\section{Annual and Semi-annual SSA Pressure Components and Variations in Polar Motion: Time Analysis}
\begin{figure}[htb]
    \centering
		\centerline{\includegraphics[width=\columnwidth]{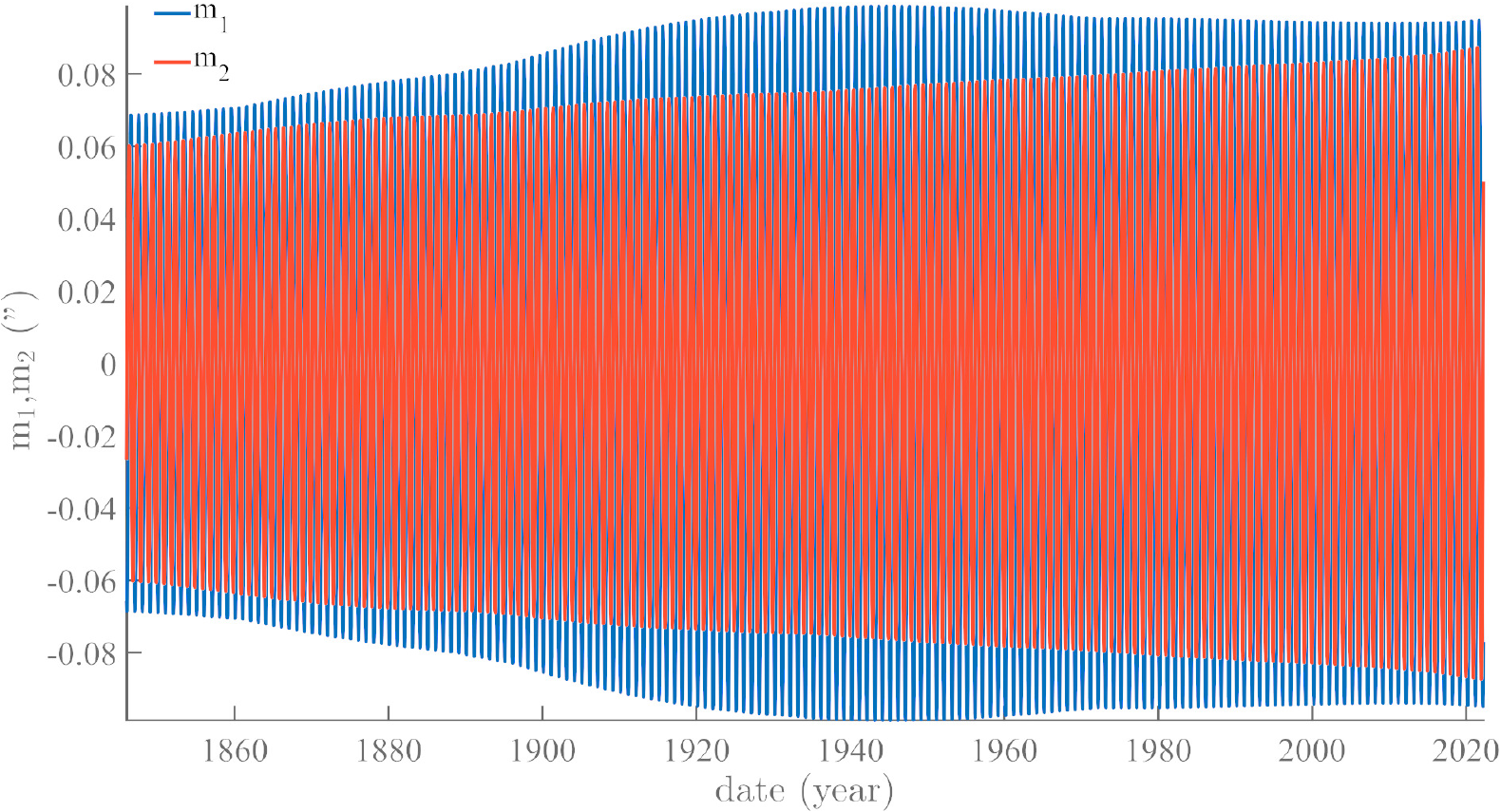}} 	
		\caption{\textbf{iSSA} annual component of rotation pole coordinates $m_1$ and $m_2$ from 1850 to the Present.}
		\label{Fig:01}
\end{figure}

\begin{figure}[htb]
    \centering
	\begin{subfigure}[b]{\columnwidth}
		\centerline{\includegraphics[width=\columnwidth]{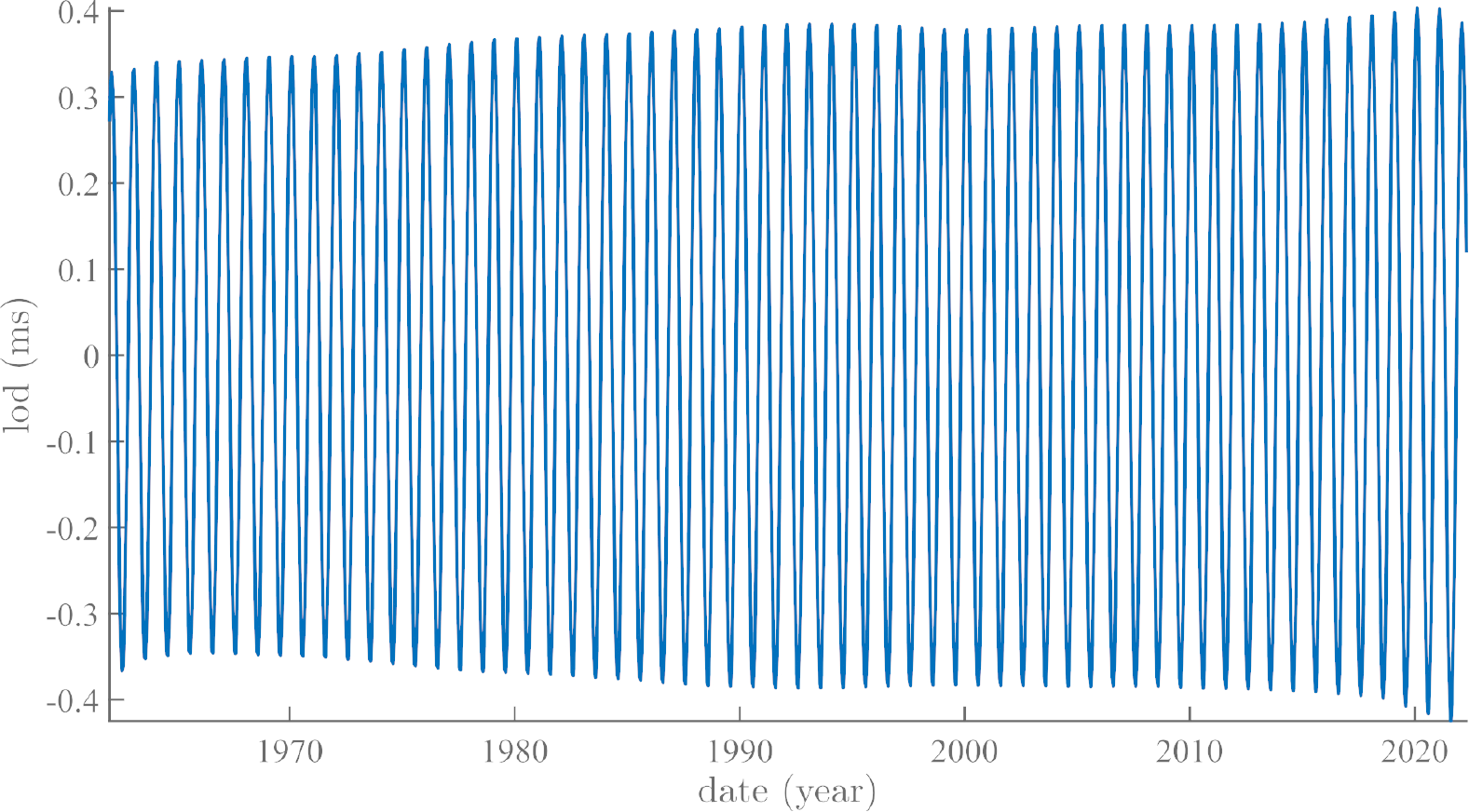}} 	
		\subcaption{\textbf{iSSA} annual component of length of day \textbf{lod} from 1850 to the Present}
		\label{Fig:02a}
	\end{subfigure}
	\begin{subfigure}[b]{\columnwidth}
		\centerline{\includegraphics[width=\columnwidth]{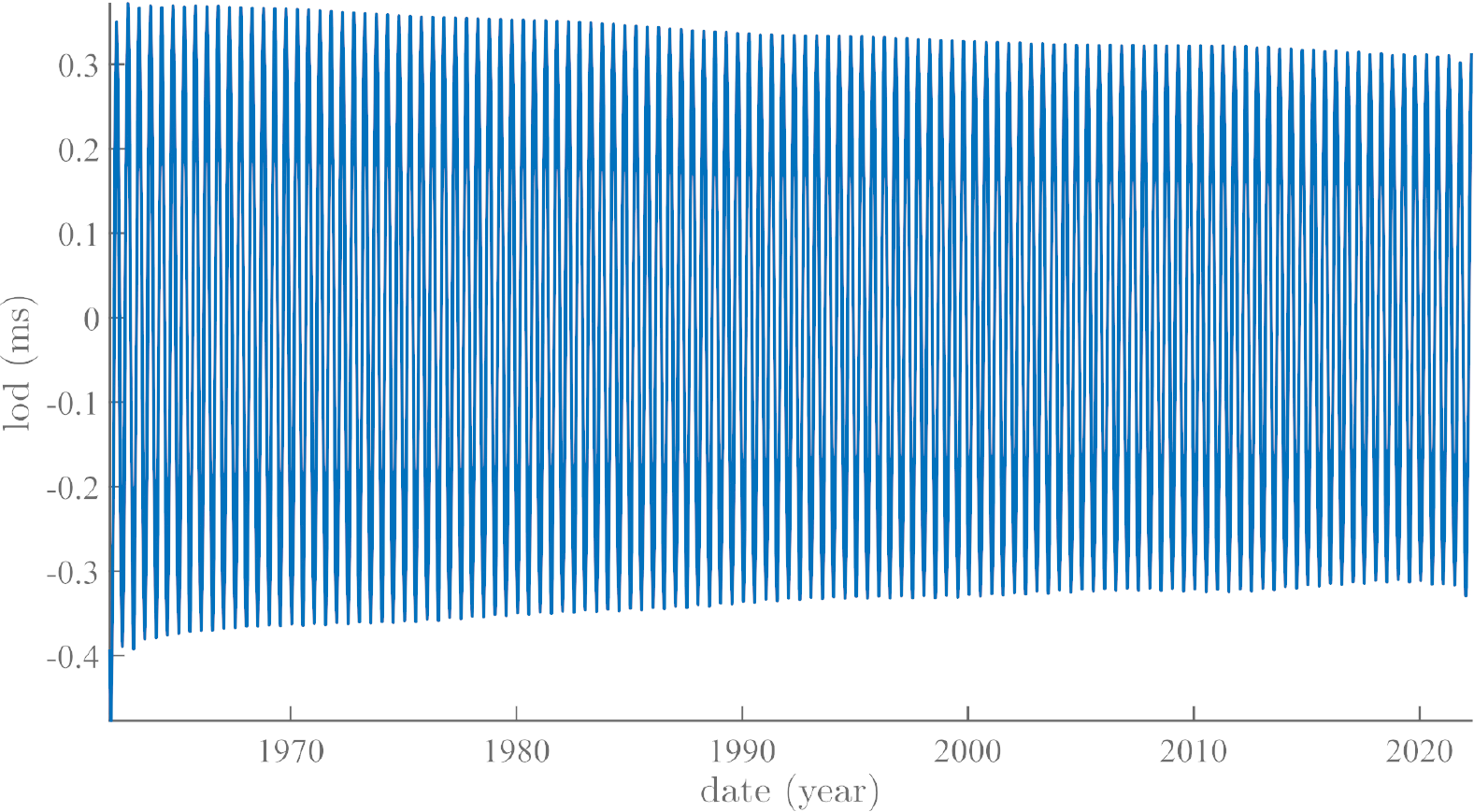}} 	
		\subcaption{\textbf{iSSA} semi-annual component of length of day \textbf{lod} from 1850 to the Present}
		\label{Fig:02b}
	\end{subfigure}
	\caption{Annual and semi-annual components extracted from length of day}
\end{figure}

    Many studies (\textit{e.g.} \cite{Jeffreys1916, Lambeck2005, Adhikari2016}) attempt to relate forced (annual) oscillations to polar motion \textbf{RP} ($m_1$, $m_2$) or length of day (\textbf{lod}) (data file EOPC01IAU2000, \cite{Vondrak1995}, maintained by IERS\footnote{ \scriptsize https://www.iers.org/IERS/EN/DataProducts/EarthOrientationData/eop.html}). The data are shown and discussed in \cite{Lopes2021}.  The annual \textbf{SSA} components of $m_1$ and $m_2$ are displayed in \Figure \ref{Fig:01}. Both undergo some amount of modulation. That modulation is significant for $m_1$, particularly between 1850 and 1930. For $m_2$, it is smaller and more regular. A $m_1$ vs $m_2$ diagram (see {\color{blue}{Appendix A}}) shows the classical Lissajou elliptical shape of annual polar motion. Note that the date of 1930 is when the Chandler wobble undergoes a phase jump of  $\pi$ (\textit{e.g.} \cite{Gibert1998, Gibert2008}). The \textbf{lod} data start only in 1962 (not shown, see \cite{Lopes2022a}). The annual and semi-annual \textbf{SSA} components of \textbf{lod} are modulated (\Figure \ref{Fig:02a} and \Figure \ref{Fig:02b}). The annual component grows slightly until 1990, flattens until 2013 and then starts growing again. The semi-annual component decreases regularly over the sixty years of available data.

\begin{figure}[htb]
    \centering
		\centerline{\includegraphics[width=\columnwidth]{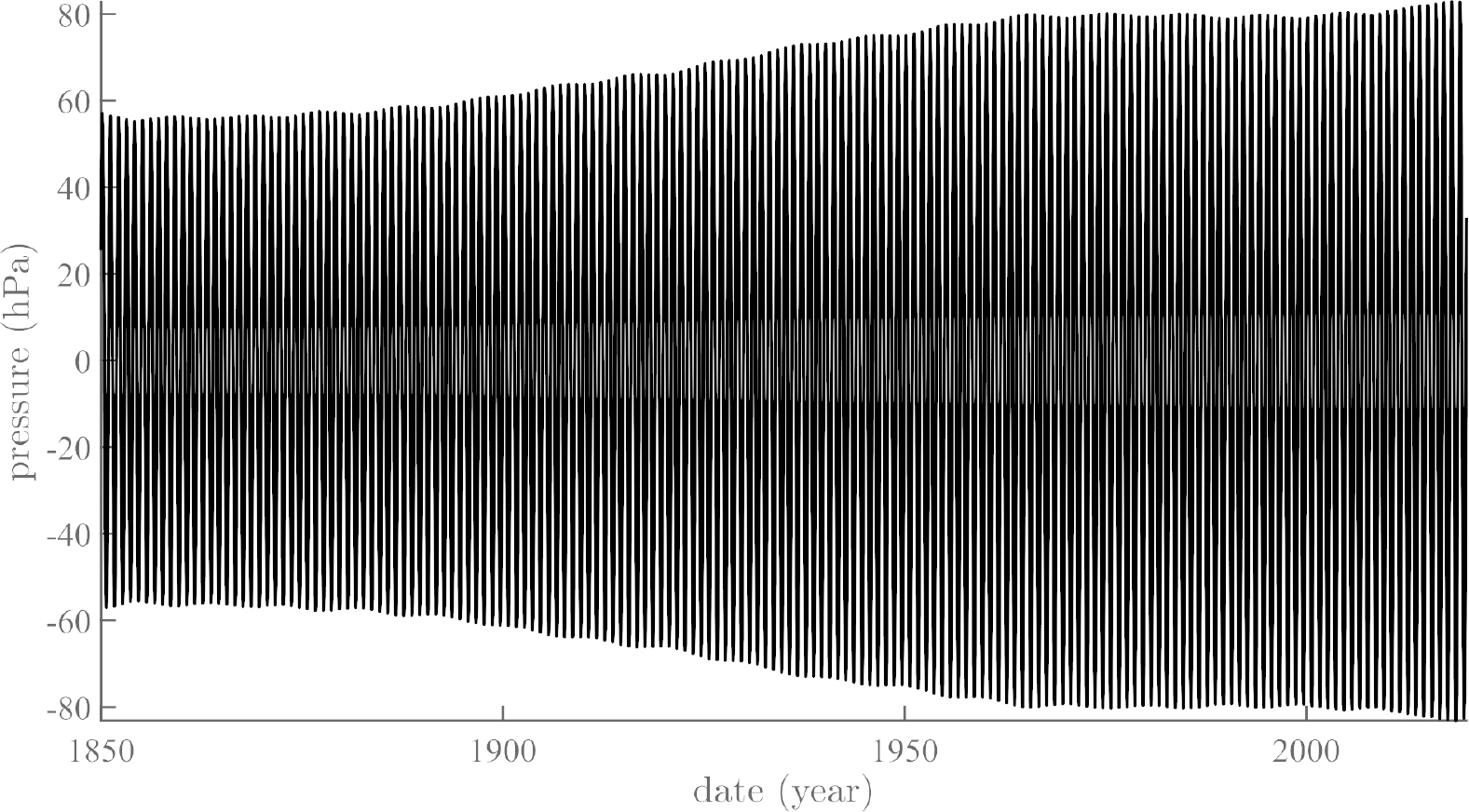}} 	
		\caption{\textbf{iSSA} annual component of sea level pressure \textbf{SLP} from 1850 to the Present.}
		\label{Fig:03}
\end{figure}

\begin{figure}[htb]
    \centering
	\begin{subfigure}[b]{\columnwidth}
		\centerline{\includegraphics[width=\columnwidth]{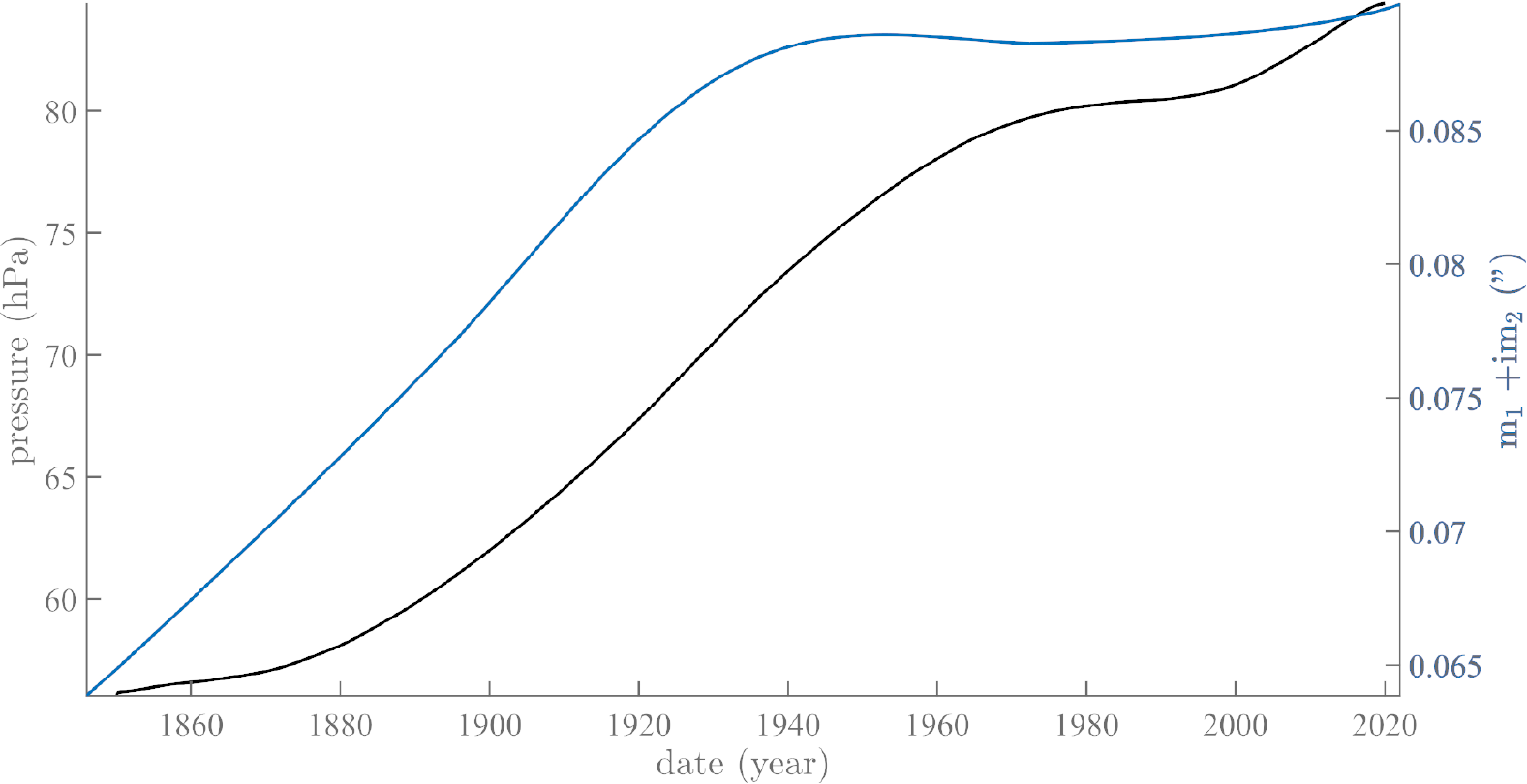}} 	
		\subcaption{Envelopes of oscillations of \textbf{iSSA} annual components of polar motion \textit{m} (blue curve, right scale) and global sea-level pressure \textbf{SLP} (black curve, left scale).}
		\label{Fig:04a}
	\end{subfigure}
	\begin{subfigure}[b]{\columnwidth}
		\centerline{\includegraphics[width=\columnwidth]{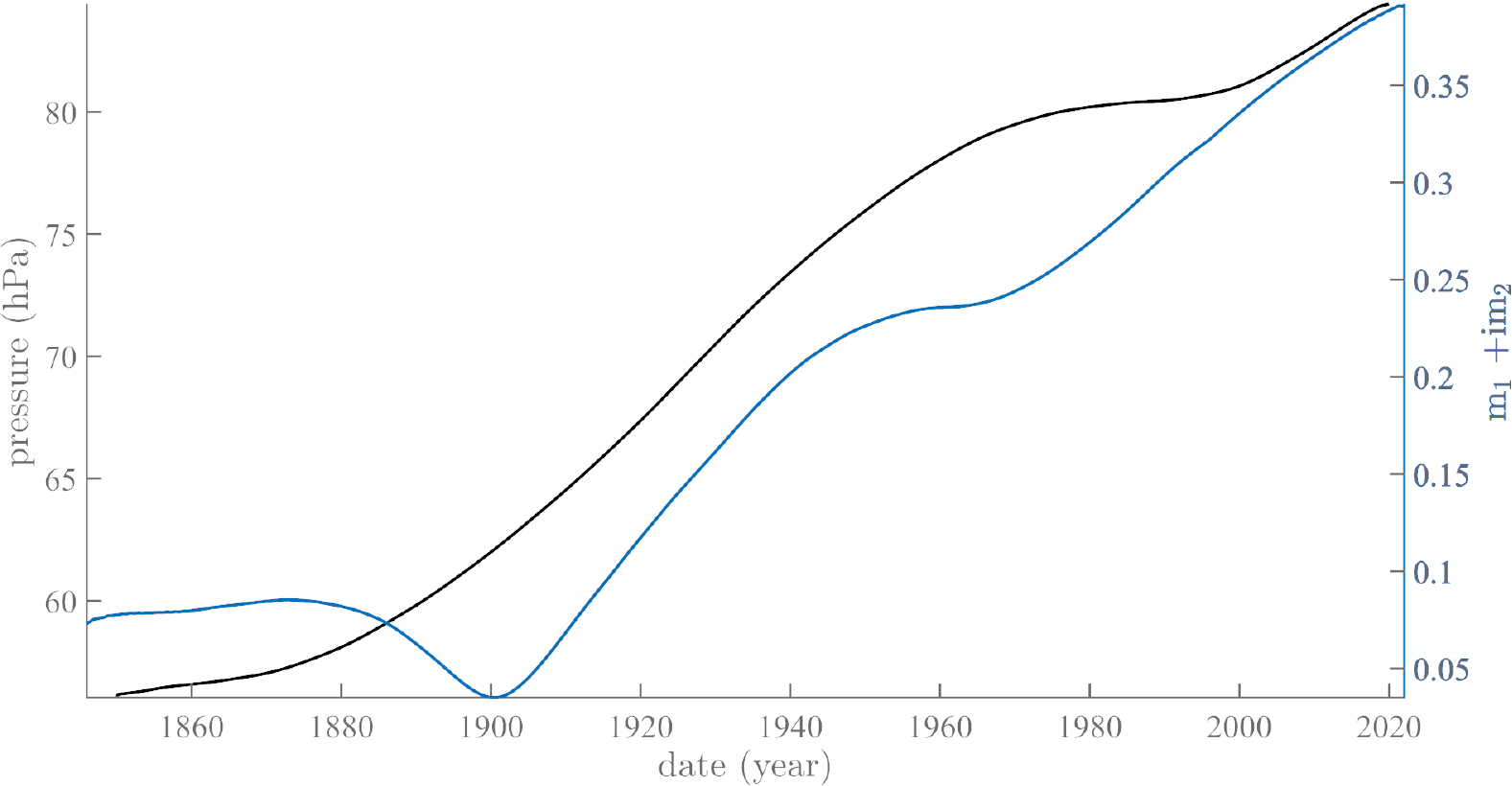}} 	
		\subcaption{Trends of the pole movement (blue curve) and atmospheric pressure \textbf{SLP} (black curve) extracted by \textbf{iSSA}.}
		\label{Fig:04b}
	\end{subfigure}
	\caption{Annual enveloppes  and trends extracted from pole movement and \textbf{SLP}}
\end{figure}

    \Figure \ref{Fig:03} shows the annual \textbf{iSSA} component since 1850, extracted from the \textbf{SLP} pressure series. Its modulation looks very much like that of $m_2$, with a phase shift (\Figure \ref{Fig:01}). We have checked that the phase shift is constant over the 70 years of data. Its mean value is 152.31 $\pm$ 2.68 days. We have already obtained (exactly) such a phase shift and note that it is (exactly) half of the Euler period of 306 days: a reason for this remains to be found. Using the Hilbert transform, we have determined the envelopes of the two annual components of polar motion and atmospheric pressure (\Figure \ref{Fig:04a}). The polar motion clearly precedes pressure by at least 40 years. On \Figure \ref{Fig:04b}, we see that the trend of polar motion is close to the envelope of pressure, in general preceding it slightly. Given the mathematical properties of the Liouville-Euler set of equations, the potential direction of causality between the two phenomena is in the sense of polar motion leading atmospheric pressure. If correct, this is an important result that confirms Laplace’s statements in his 1799 Traité de Mécanique Céleste (volume 5, chapter 1, page 347; reproduced in {\color{blue}{Appendix B}}). \\
    
Thus, after applying the Kepler’s law of conservation of areas, Laplace concludes that atmospheric motions do not affect pole rotation.

\section{The Annual and Semi-annual SSA Pressure Components: Spatial Analysis}    
    We show in \Figure {\color{blue}{5}} polar maps of the mean amplitude of the annual component (oscillation), for each season since 1850, and in \Figure {\color{blue}{6}}  polar maps of the mean of the semi-annual component. \\
    
     In \Figure {\color{blue}{5}}, in the Spring \textbf{SH}, there is a very regular pattern at the intersections of the 20$^{\circ}$S  to 50$^{\circ}$S annulus and the southern tips of the three southern continents (Africa, Australia and South America). There is a smaller dipole with its negative part centered almost perfectly on the South pole, and its positive part centered on 70$^{\circ}$S, 105$^{\circ}$W. In the Fall, the pattern is the same but with a reversal in sign. In the Summer, there is a positive band between 15 and 25°S latitude from Madagascar to New Caledonia (40°-170°E) and a small negative spot near the South pole (Winter is the same with a sign reversal). The positive annulus is actually also seen, much weaker, in the Atlantic and Pacific oceans. The \textbf{NH} is rather different with, in the Spring, a large negative area over Asia from the Red Sea to Japan and extending in latitude from 80$^{\circ}$N to 15$^{\circ}$N. This large “anomaly” (using geophysical language) extends negative arms towards Senegal and over the western USA. Over the year, the “anomaly” changes sign. Still in the Summer, there are two sizeable positive anomalies over the northern Atlantic and Pacific Oceans between 40$^{\circ}$N and 60$^{\circ}$N. In the \textbf{SH}, the large mean annual pressure variations occur over the Southern continents; in the \textbf{NH}, the Asian continental “anomaly” dominates, though two significant features occur over the northern parts of the oceans. Because the Asian “anomaly” and the two (opposite sign) anomalies in the North Atlantic and Pacific are separated by almost 180° in longitude, their maxima being always in phase, their contributions to the seasonal excitation function tend to cancel \citep{Lamb1972}. \\

    The semi-annual component (\Figure {\color{blue}{6}}) has stronger patterns of symmetry. The \textbf{SH} features a very stable 3-fold symmetry (triskel); three large anomalies form an equilateral triangle centered on 70$^{\circ}$E, 30$^{\circ}$W and 210$^{\circ}$W, between latitudes 40$^{\circ}$S and 60$^{\circ}$S, a large anomaly of opposite sign is centered on the South Pole (the maximum is actually offset from the \textbf{SP} by 20$^{\circ}$E towards the 120$^{\circ}$E meridian), and an anomaly with the same sign as that over the pole, but located over southern Africa. The \textbf{NH} has weaker anomalies forming an irregular triangle with apices near the North Pacific (40$^{\circ}$N, 165$^{\circ}$E), Baffin Sea (40$^{\circ}$N, 50$^{\circ}$W) and Iran (30$^{\circ}$N, 60$^{\circ}$W). \\
    
        The bottom two maps in \Figure {\color{blue}{7}} are the Spring maps for the annual and semi-annual components. We can compare the maps of Figures {\color{blue}{5}}, {\color{blue}{6}}   and {\color{blue}{7}} for the annual, semi-annual and trend components with corresponding figures from the reference treatise of \cite{Lamb1972}, reproduced in {\color{blue}{Appendix C}} .Recall that taken together these three components suffice to capture about 85\% of the total signal variance. Figures 4-12 and 4-13 for instance (reproduced in {\color{blue}{Appendix C}} as \Figures  {\color{blue}{C1a}}and  \Figure {\color{blue}{C1b}}) show Lamb's representation of the average mean sea level pressure over (respectively) the northern and southern hemispheres in the 1950s. There is very good agreement for the northern hemisphere (\Figure \ref{Fig:C1a}): the January map from Lamb and our Winter map show the dominant “dipole” with the large high pressure (HP) over Asia and the low pressure (LP) over the northern Pacific. The April and Spring maps do not agree so well. The July map from Lamb and our Summer map both show a large LP over Asia and two HPs over the northern Pacific and northern Atlantic. The October map from Lamb and our Autumn map show a weaker HP over Asia. For the southern hemisphere (\Figure \ref{Fig:C1b}), agreement between the two sets of maps is significantly less; the Winter vs January maps both feature a HP centered on Antarctica and a LP extending from Madagascar to Australia, and the reverse in Summer/July. Our Spring (respectively Autumn) maps feature strong HPs (respectively LPs) on the southern tips of the southern continents. This sharp pattern is not really seen so well in {\color{blue}{Lamb’s (1972)}} maps ({\color{blue}{Appendix C}}). \\
        
     Lamb’s Figure 3-17 (reproduced in {\color{blue}{Appendix C}} as \Figure {\color{blue}{C2}}) represents the annual mean distribution of atmospheric pressure at sea level for the time span 1900-40 for the northern hemisphere and 1900-50s for the southern hemisphere. This can be compared to our maps of the trend (\cite{Lopes2022b}, and \Figure {\color{blue}{7}}, top row). The agreement is quite good: the 3 fold symmetry of the northern hemisphere and the 4 fold symmetry of the southern hemisphere are even clearer in the \cite{Lopes2022b} maps using the \textbf{iSSA} method to extract the trend (i.e. the main component) from the data. As concluded by \cite{Lamb1972} the effects of geography modify the circumpolar circulation much more in the northern than in the southern hemisphere; “ The broad zonal character of the long-period mean circulation is as permanent as the circumpolar arrangement of the heating and cooling zones and the circumpolar vortex aloft”. It appears to us that such maps are not available in the published literature; the maps from the southern hemisphere are particularly interesting, and so is the triskel in \Figure {\color{blue}{7}} (top left). Specialists may wish to analyze these maps in more detail; they can be provided on simple request. 
     \newpage
    
\begin{figure}[H]
    \centering
	\begin{subfigure}{\columnwidth}
		\centerline{\includegraphics[width=\columnwidth]{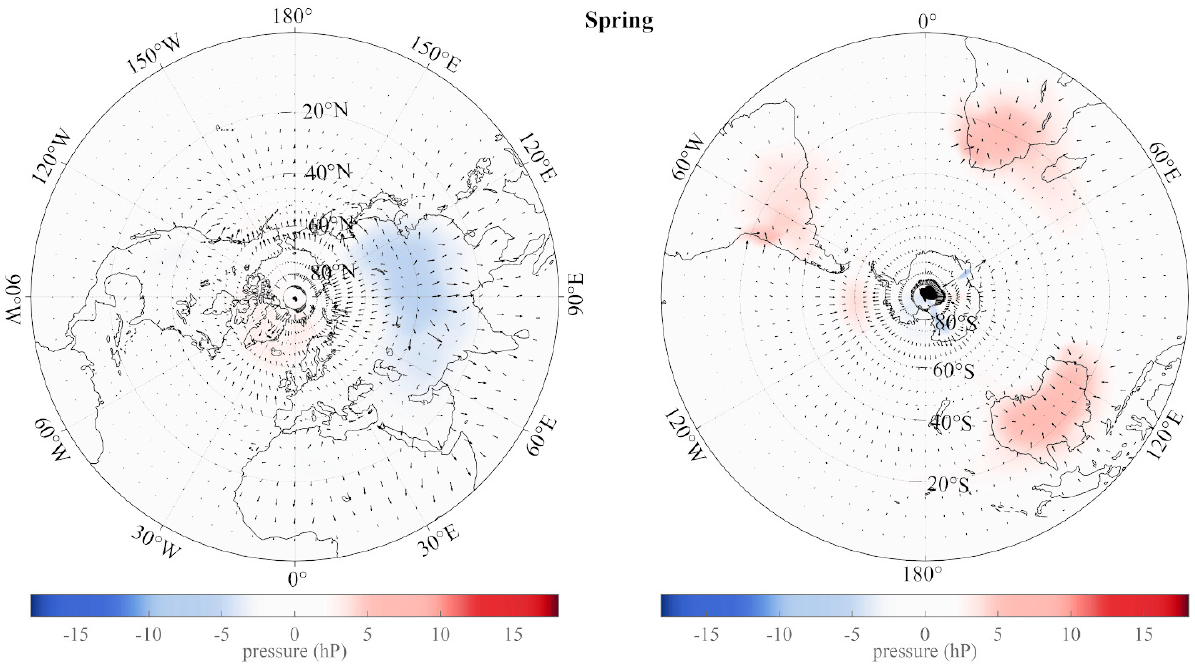}} 	
		\subcaption{During springs}
	\end{subfigure}
	\begin{subfigure}{\columnwidth}
		\centerline{\includegraphics[width=\columnwidth]{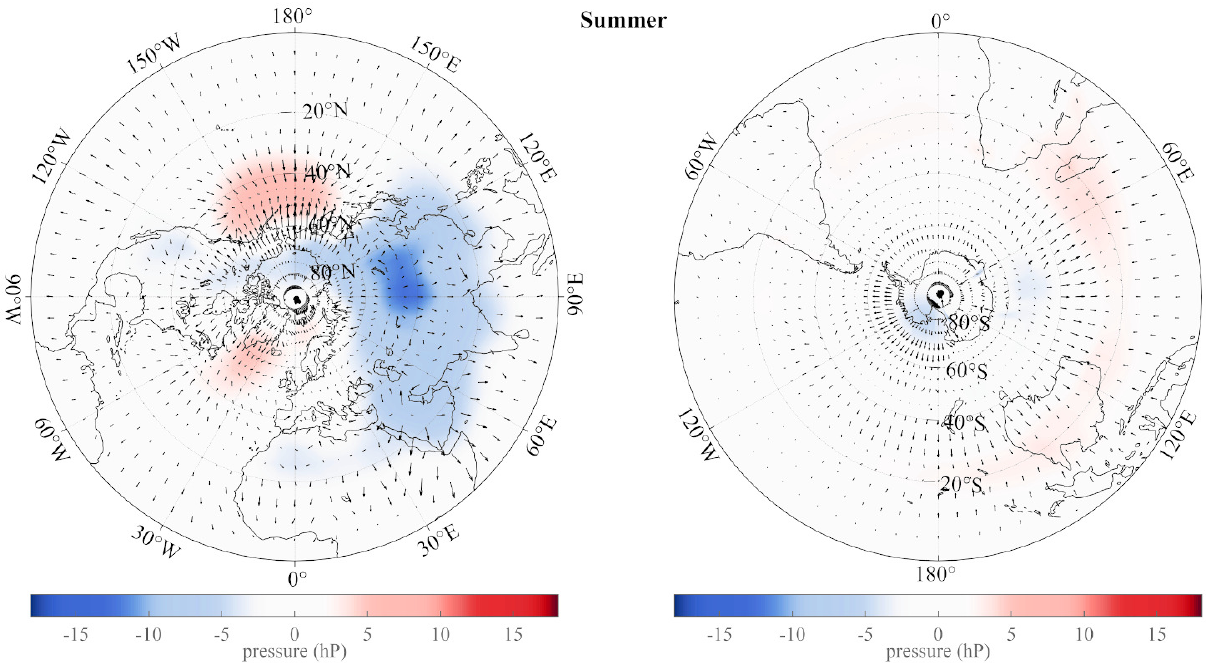}} 	
		\subcaption{During summers}
	\end{subfigure}
	\begin{subfigure}{\columnwidth}
		\centerline{\includegraphics[width=\columnwidth]{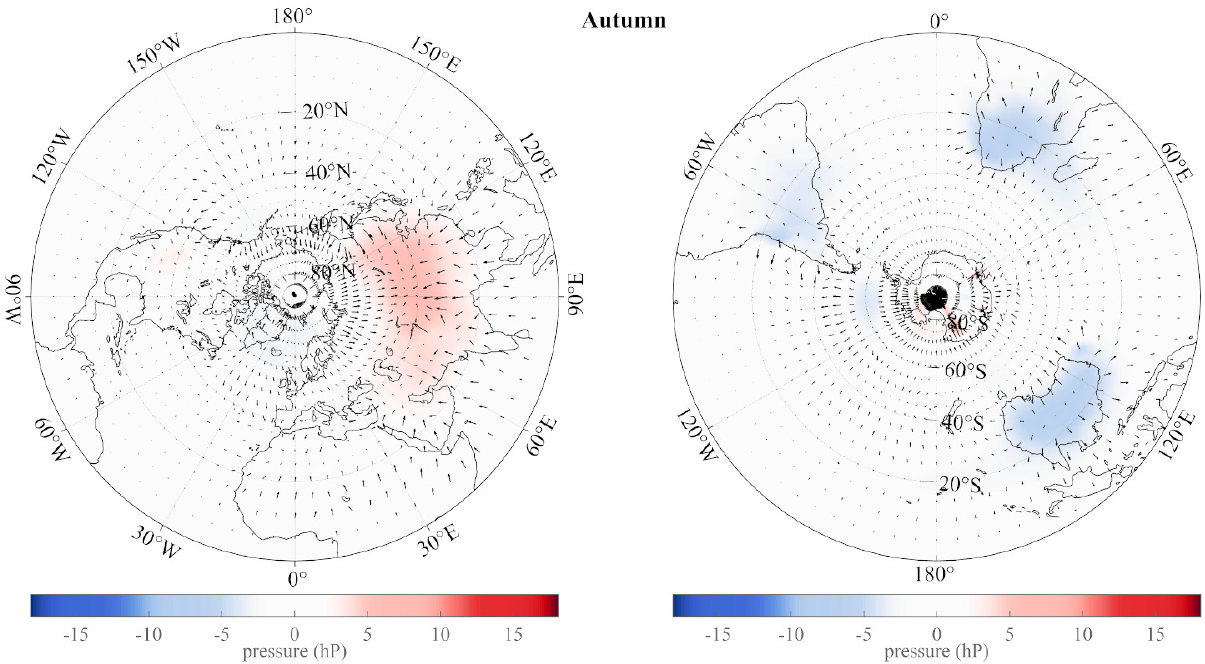}} 	
		\subcaption{During autumns}
	\end{subfigure}
	\begin{subfigure}{\columnwidth}
		\centerline{\includegraphics[width=\columnwidth]{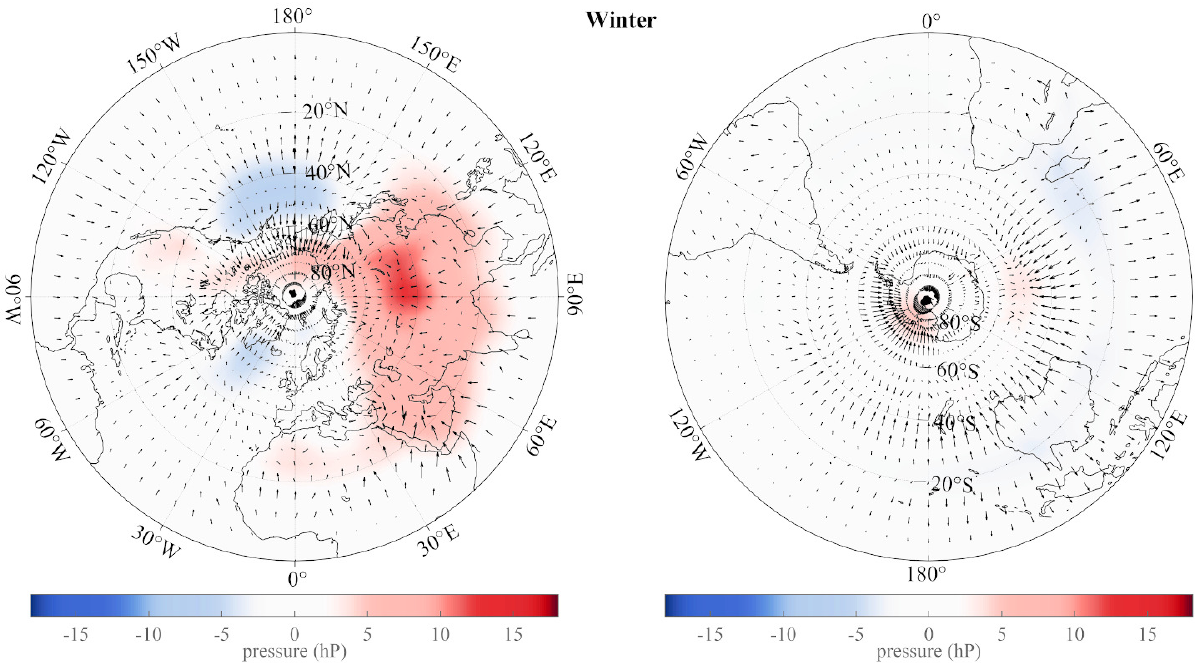}} 	
		\subcaption{During winters}
	\end{subfigure}
	\label{Fig:05}
	\caption{Polar views of North hemisphere (left column) and South hemisphere (right column) showing the mean (1850-2020) of the annual oscillation (\textbf{iSSA} Component 2) for all (from top to bottom) Springs, Summers, Autumns and Winters since 1850.}
\end{figure}	
    
\begin{figure}[H]
    \centering
	\begin{subfigure}[b]{\columnwidth}
		\centerline{\includegraphics[width=\columnwidth]{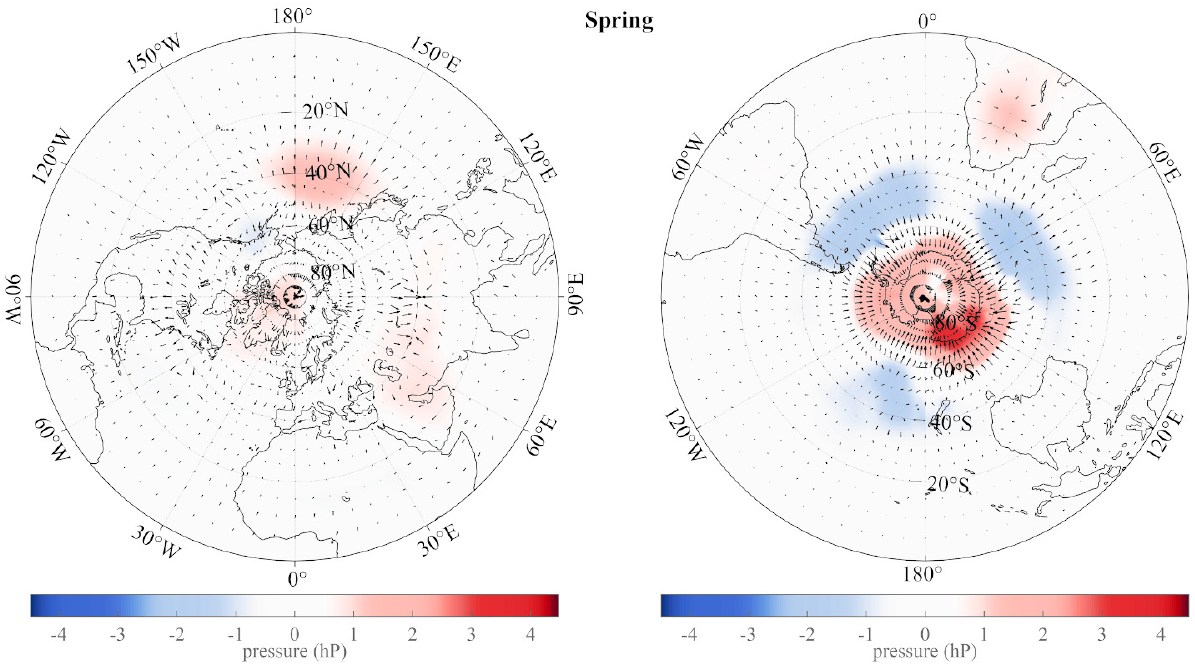}} 	
		\subcaption{During springs}
	\end{subfigure}
	\begin{subfigure}[b]{\columnwidth}
		\centerline{\includegraphics[width=\columnwidth]{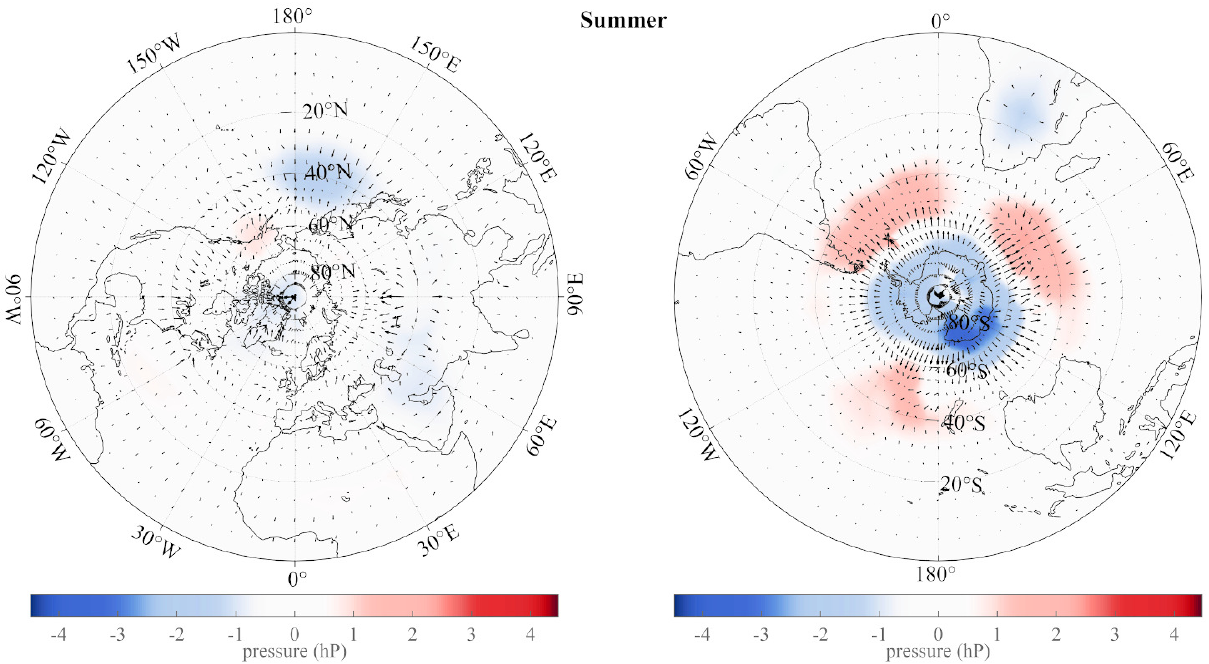}} 	
		\subcaption{During summers}
	\end{subfigure}
	\begin{subfigure}[b]{\columnwidth}
		\centerline{\includegraphics[width=\columnwidth]{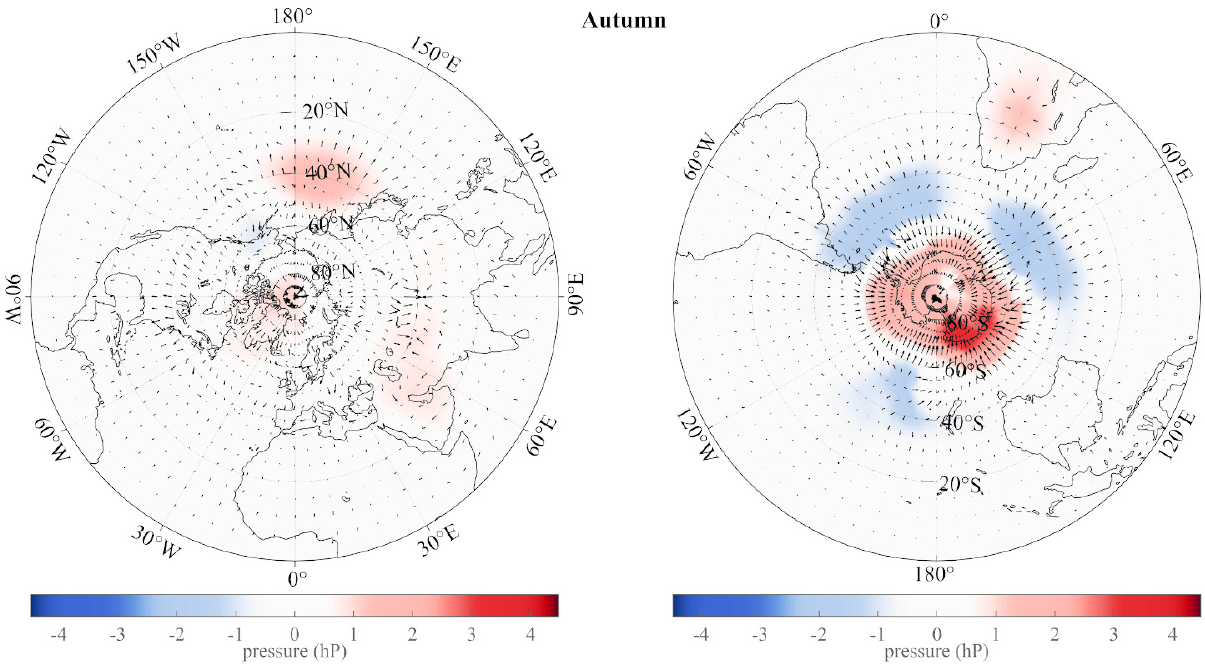}} 	
		\subcaption{During autumns}
	\end{subfigure}
	\begin{subfigure}[b]{\columnwidth}
		\centerline{\includegraphics[width=\columnwidth]{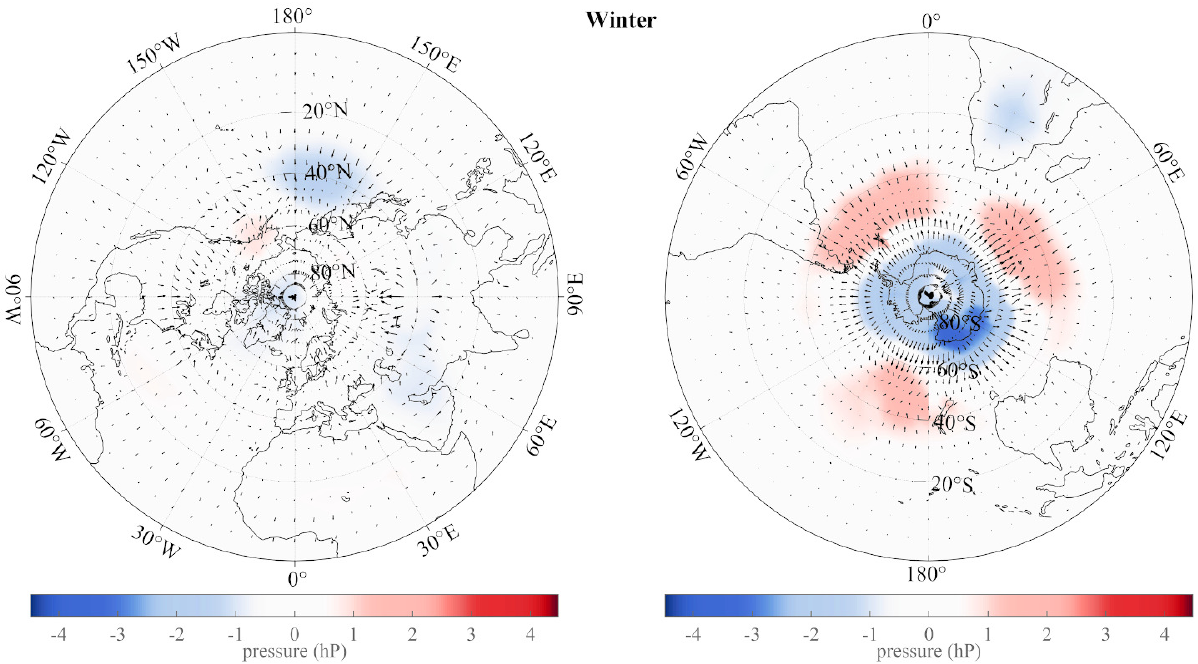}} 	
		\subcaption{During winters}
	\end{subfigure}
	\label{Fig:06}
	\caption{Polar views of North hemisphere (left column) and South hemisphere (right column) showing the mean (1850-2020) of the semi-annual oscillation (\textbf{iSSA} Component 3) for all (from top to bottom) Springs, Summers, Autumns and Winters since 1850.}
\end{figure}

\begin{figure}[H]
    \centering
	\begin{subfigure}[b]{\columnwidth}
		\centerline{\includegraphics[width=\columnwidth]{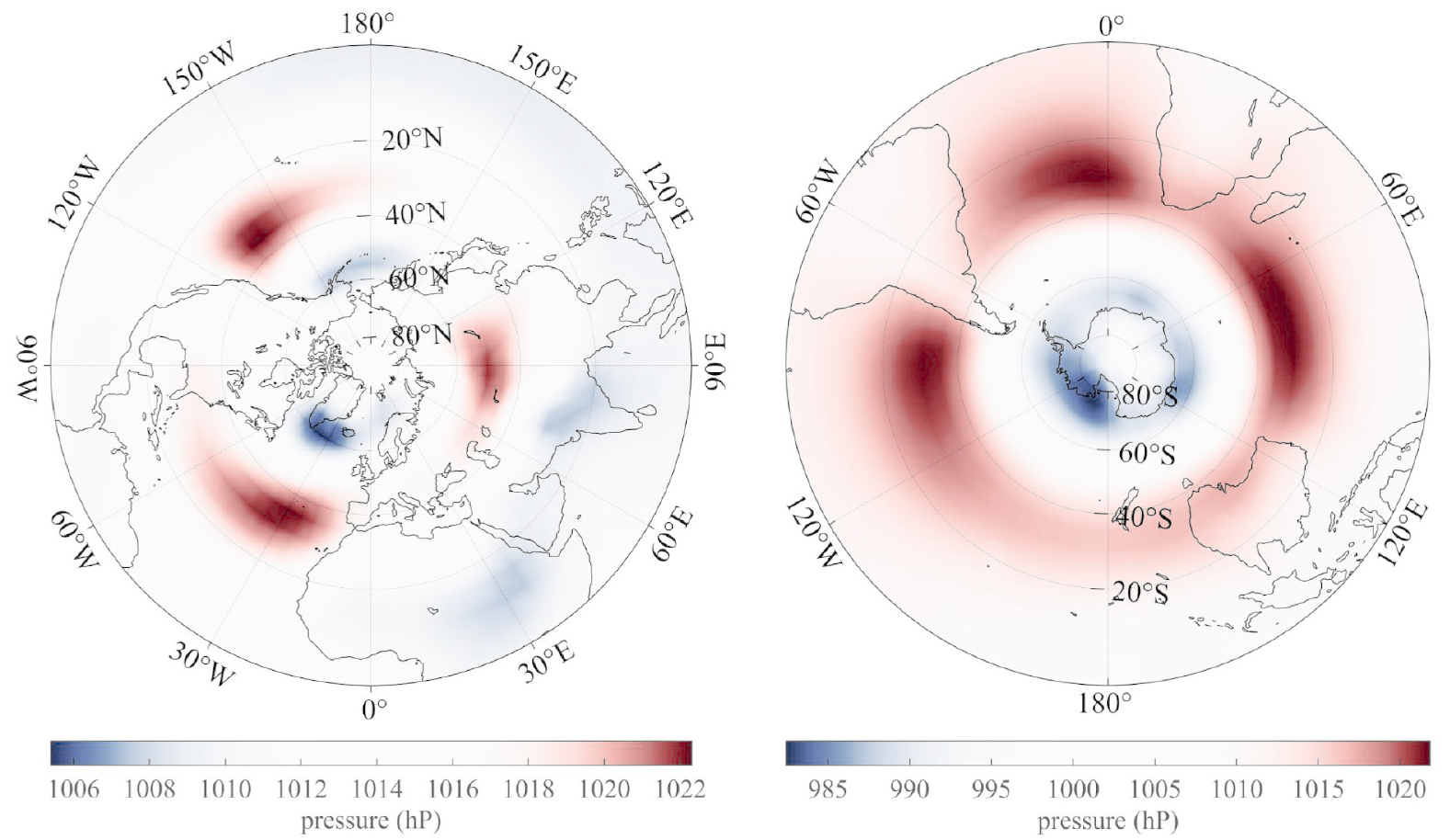}} 	
	\end{subfigure}
	\begin{subfigure}[b]{\columnwidth}
		\centerline{\includegraphics[width=\columnwidth]{figure_arxiv/fig_05a.pdf}} 	
	\end{subfigure}
	\begin{subfigure}[b]{\columnwidth}
		\centerline{\includegraphics[width=\columnwidth]{figure_arxiv/fig_06b.pdf}} 	
	\end{subfigure}
	\label{Fig:07}
	\caption{Polar views of North hemisphere (left column) and South hemisphere (right column) showing the trend (iSSA component 1 from \cite{Lopes2022b}, and maps of \textbf{iSSA}components 2 (annual) and 3 (semi-annual) for all Springs since 1850.}
\end{figure}	    
               
\section{Discussion}
    \subsection{Symmetries and Forcings}
    We pointed out in a preliminary paper on symmetries of global sea level pressure \citep{Lopes2022b} that the pattern of \Figure {\color{blue}{7}} was remarkably stable over the 150 years of the data, and that the observed geometry could be modeled as Taylor-Couette flow of mode 3 (\textbf{NH}) or 4 (\textbf{SH}). The remarkable regularity and order three symmetry of the \textbf{NH} triskel occurs despite the lack of cylindrical symmetry of the northern continents. The stronger intensity and larger size of features in the \textbf{SH} is linked to the presence of the annular currents. Following Kepler and Laplace (see above) it appears that the molecules in the atmosphere cannot perturb polar axis motion (“\textit{Nous sommes donc assurés qu’en même temps que les vents analysés diminuent ce mouvement, les autres mouvements de l’atmosphère qui ont lieu au-delà des tropiques, l’accélèrent de la même quantité} “). The atmosphere behaves as a rotating cylinder undergoing stationary flow. The top of the cylinder is a free surface, the bottom is the surface of the solid crust and/or fluid ocean. This topography interacts strongly with the atmosphere.
    
    \subsection{Sun-Earth Distance and Phases}
    The seasonal change in sign agrees with changes in the Sun-Earth distance. The present values of the ellipticity of the Earth’s orbit and the obliquity of its rotation axis lead to changes of orbital velocity with Summer in the southern hemisphere at perihelion and Summer in the northern hemisphere at aphelion. The torque associated with the Sun changes as velocity changes. As discussed in \cite{Lopes2021,Lopes2022a}, forces acting on the polar axis do so through a torque ($m_1$,$ m_2$) obeying the Liouville-Euler system of equations (from {\textcolor{blue}{Laplace, 1799}}). The influence of these torques is in quadrature with the length of day (\textbf{lod}). Changes in orbital velocity of our planet lead to a variation in the solar torque, which acts against the sense of rotation of the planet. Thus, annual variations in lod should be in opposition to changes in the Sun-Earth distance, and they are (\Figure {\color{blue}{8a}}, bottom). In the northern latitudes, both the Earth's rotation and the winds go from West to East, the Earth’s rotation slows in Summer and accelerates in Winter, hence positive pressure anomalies in Winter, negative in Summer. In other words, the pressure anomalies are due to the variation in rotation velocity (tracked through lod changes and winds). The phase difference between the annual components of lod and \textbf{SLP} is constant at 32 $\pm$ 1.2 days (\Figure  {\color{blue}{8b}}). Relative variations of amplitudes should be similar and indeed they are (\Figure {\color{blue}{8b}}). 
    
    \newpage
\begin{figure}[H]
    \centering
	\begin{subfigure}[b]{\columnwidth}
		\centerline{\includegraphics[width=\columnwidth]{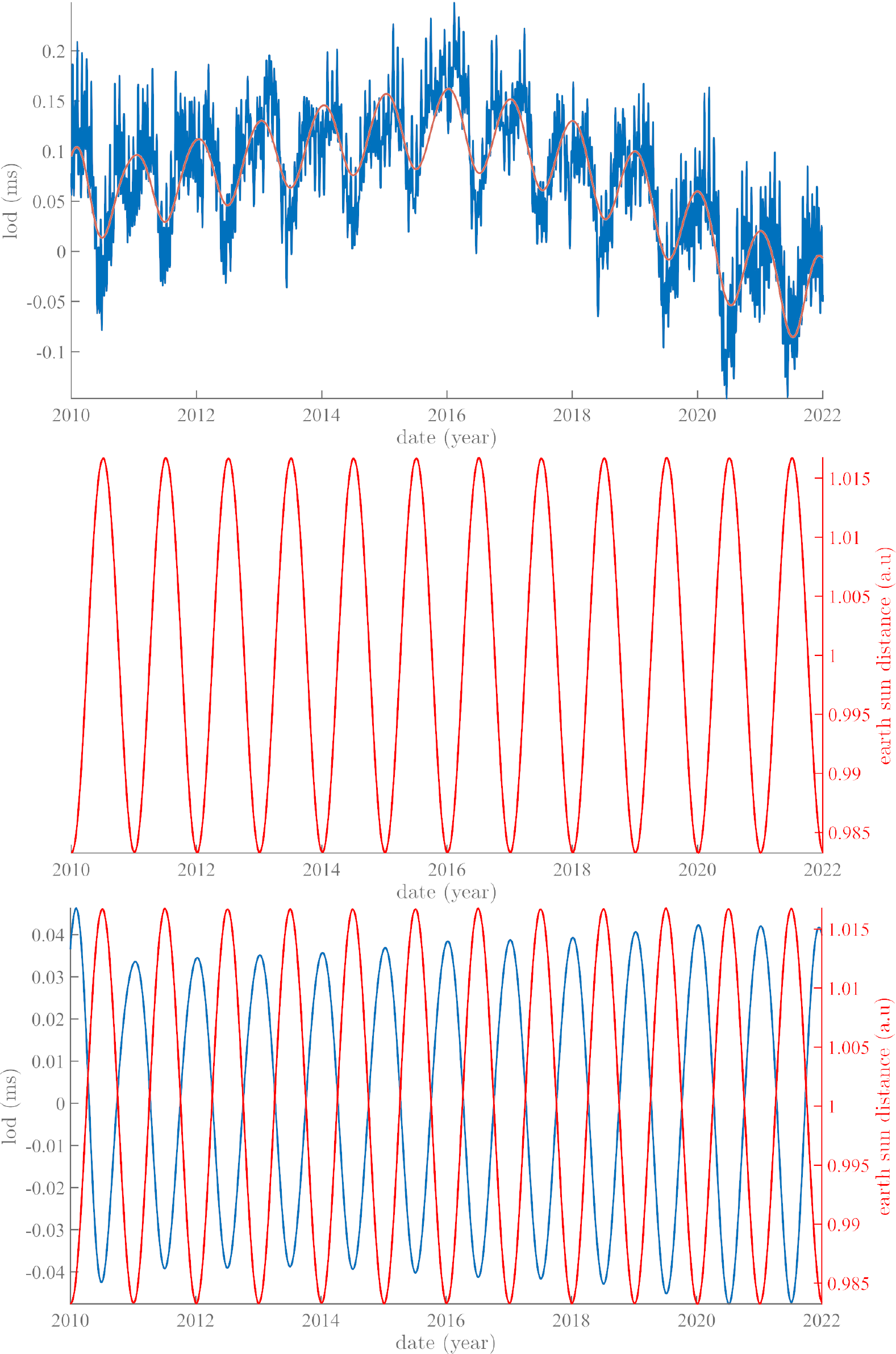}} 	
     	\subcaption{(top) \textbf{lod} data (blue) and their annual component (red); (middle) variation in the Sun-Earth distance (red); (bottom) the two curves above superimposed, showing phase opposition.}
		\label{Fig:08b} 	
	\end{subfigure}
	\begin{subfigure}[b]{\columnwidth}
		\centerline{\includegraphics[width=\columnwidth]{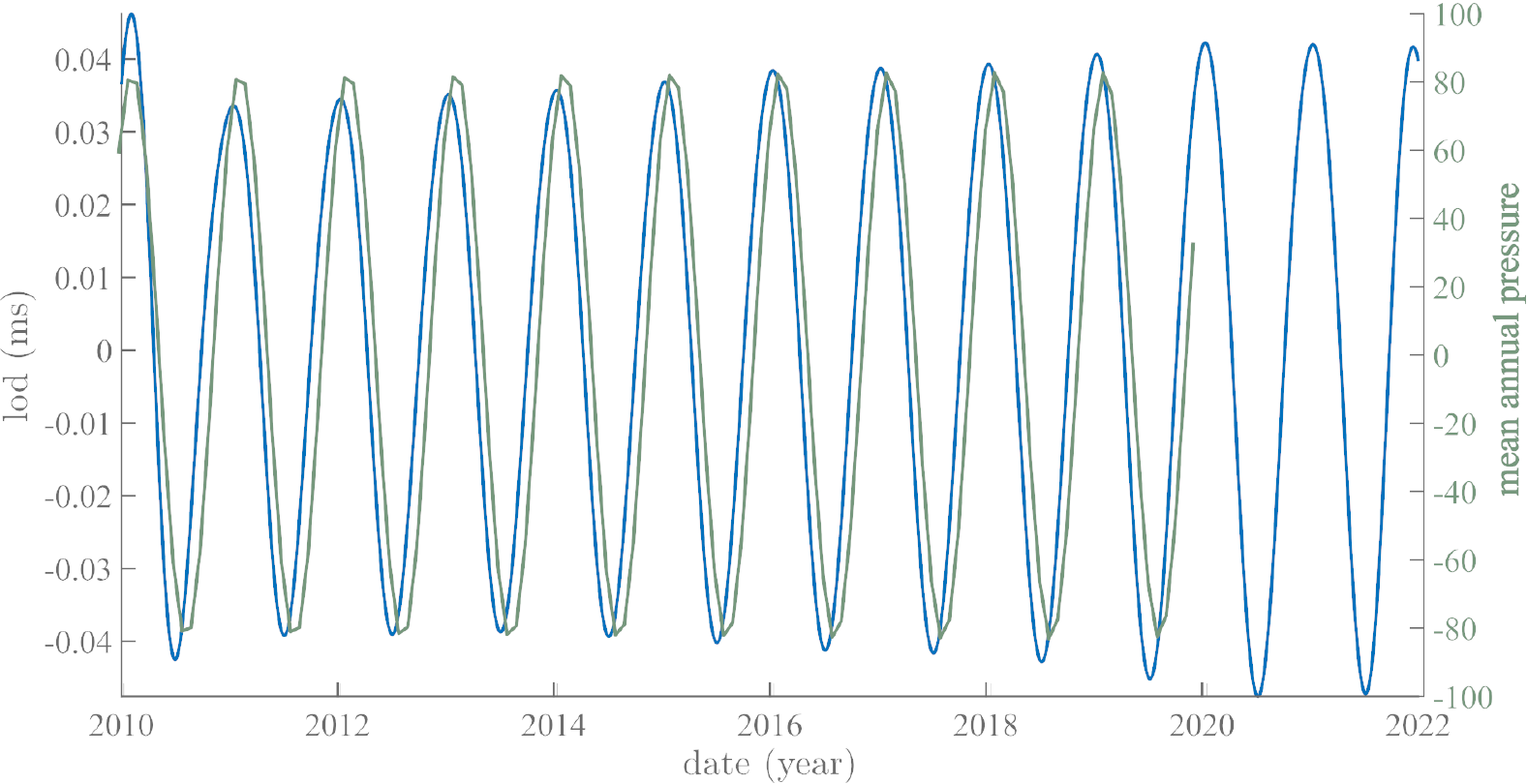}}
		\subcaption{In green the annual \textbf{SSA} component 2 of global sea level pressure, and in blue the annual component of \textbf{lod}. The phase difference is constant at 32 $\pm$ 1.2 days.}
		\label{Fig:08b} 	
	\end{subfigure}
	\label{Fig:08}
	\caption{Astronomicals and geophysicals annual oscillations components}
\end{figure}	    

\section{Sketch of a Mechanism}
    Because more than 70\% of the total signal variance is captured by the trend of the atmospheric pressure, that varies by only 1 per mil since 1850, one can resort to the theory of turbulent flow in the case of infinitesimal perturbations (\textit{cf.} {\color{blue}{Landau and Lifchitz, 1987}}). The non stationary $v_1(r,t)$ perturbation of the stationary solution $v_0(r)$ verifies the Navier-Stokes equation for an incompressible fluid:
 \begin{equation}
     \dfrac{\partial v_{1}}{\partial t} + (v_{0} \nabla)v_{1} + (v_{1} \nabla) v_{0}= -\dfrac{p_{1}}{\rho} + \nu.\mathrm{grad} v_{1}, \quad \mathrm{div} v_{1} = 0
 \end{equation}

The general solution can be written as the sum of particular solutions where the dependence of $v_1$ on time is of the form $e^{-i\omega t}$. There is still no theoretical basis for a mathematical solution of the stability of flow about finite dimension bodies plunged in a rotating fluid (\textit{eg.} \cite{Chandrasekhar1961, Landau1987,Frisch1995}). Only experiments show that flow stability depends on the relative value of the Reynolds number ($R$), each type of flow having its critical Reynolds number ($R_c$), beyond which instability sets in (\textit{cf.} \cite{Landau1944}. It is still difficult to solve turbulent flows other than in the case of concentric cylinders (\textit{e.g.} \cite{Schrauf1986, Mamun1995, Nakabayashi1995, Hollerbach2006, Mahloul2016,Garcia2019,Mannix2021}). But fortunately, this is useful in the case of atmospheric and oceanic flows on the spherical Earth, as is suggested (among other evidence) by the annular structures shown in \Figure {\color{blue}{5}} to \Figure {\color{blue}{7}}. \\

    \cite{Rayleigh1916} tackled the study of stability of a fluid comprised between two concentric rotating cylinders at high $R$. It was generalized for any value of $R$ by \cite{Taylor1923} who showed that particular independent solutions were given by:  
\begin{equation}
    v_1(r,\varphi,z) = f(r)*e^{i(n\varphi + kz - \omega t)}
\end{equation}

$n$ being an integer ($n$=0 for axial symmetry), $k$ the real wavelength of the instability, and w an acceptable frequency solving the equations in plane $z$=constant with $v_1$=0 on cylinders with radii $r=R_1$ and $r=R_2$ (see also the paper on triskels by \cite{Lopes2022a}. The Reynolds number is evaluated by $\Omega_1 R_1^2 /\nu$ (or  $\Omega_2 R_2^2 /\nu$). For fixed $n$ and $k$, the $\omega$ become a discrete set of $\omega_n(k)$ values. In the case of the Earth, $v_1$=0 at the solid surface and at the atmosphere free surface, and the only observed solutions are 24 h and 1 yr. We therefore expect to find harmonics of these two values, The detailed study for two concentric cylinders can be found in \cite{Chandrasekhar1961}. The $v_1$ flow is stationary and consists in toric vortices (or Taylor vortices) regularly spaced along the cylinder generatrices. If the series of periods $\omega_n(k)$  linked to the annual oscillation is 1, 1/2 1/3, 1/4 yr (see section 3), then one expects to find experimentally perturbations of order $n$=3 or 4, as was already the case for trends (\cite{Lopes2022b}; \Figure {\color{blue}{8}} top). We find in the present study that indeed such is the case for the annual (\Figure {\color{blue}{7}} middle) and semi-annual (\Figure {\color{blue}{7}} bottom) oscillations.

\section{Conclusion}
    This paper has attempted to obtain better constraints on the sources of the trend and annual components of both global sea-level pressure (\textbf{SLP}) and variations in the Earth's rotation (\textbf{RP} – coordinates of the rotation pole - and \textbf{lod} --  length of day), and to test the hypothesis that there might be a direct or indirect causal link between them. In a previous study, it had been shown, using iterative singular spectrum analysis (\textbf{iSSA}), that the mean sea level atmospheric pressure (\textbf{SLP}) contains, in addition to a weak trend, a dozen quasi-periodic or periodic components ($\sim$130, 90, 50, 22, 15, 4, 1.8 yr), the annual cycle and its first three harmonics \citep{Lopes2022a}. These periods were already known to be characteristic of the space-time evolution of the Earth's rotation axis (\textbf{RP}) and had been identified in many characteristic features of solar and terrestrial physics. Polar drift and the free Chandler wobble having been studied in earlier papers \citep{LeMouel2021, Lopes2022a, Lopes2021}, there remained a need of a similar analysis of the annual and semi-annual components of \textbf{SLP}, \textbf{RP} and \textbf{lod}. Most older and many recent analyses suggested more or less tentatively that the annual oscillation in the \textbf{RP} was due to a geographical redistribution of crustal and atmospheric masses, though it was agreed that quantitative modeling of this was unsatisfactory \citep{Lambeck2005}. \\
    
    In the present paper, we have undertaken a more complete analysis of the annual \textbf{SSA} component of mean sea level global pressure \textbf{SLP}. With the use of iterative \textbf{SSA} (\textbf{iSSA}), we have extracted the components in each grid cell and as a function of time since 1850. In section 3, we have listed the components and their amplitudes (in hPa). We recognize 11 quasi-periodic components plus the trend. This trend averages 1009 hPa and varies by only 0.7 hPa over the 170 year period with available data. The amplitudes of the annual component and its 3 first harmonics decrease from 93 hPa for the annual to 21 hPa for the third harmonic. In contrast, the components with pseudo-periods longer than a year range between 0.2 and 0.5 hPa. The trend is as large as the annual (21 hPa) and could be part of the 90 yr Gleissberg cycle. \\
    
    We have taken the opportunity of this study to further analyze the components of polar motion \textbf{RP} (coordinates $m_1$ and $m_2$) and length of day (\textbf{lod}). Comparing \Figure {\color{blue}{1}} and \Figure {\color{blue}{3}}, we see that the annual components of \textbf{RP} and \textbf{SLP} have a phase difference of 152 $\pm$ 2 days, which is constant over the 70 years of common data; these components are modulated in the same way, growing in amplitude between $\sim$1870 and $\sim$1960. We note that the phase difference happens to be almost exactly half of the Euler period of 306 days, for reasons we do not know so far. There is actually a phase lag of $\sim$40 years between \textbf{RP} and \textbf{SLP}, as suggested by \Figure {\color{blue}{4a}}  (showing the envelopes of the components in \Figure {\color{blue}{1}}  and \Figure {\color{blue}{3}}) and to a lesser extent by \Figure {\color{blue}{4b}}  (showing their first components, the trends). With the Euler-Liouville system of equations {\color{blue}{Laplace,1799;  Lopes et al. 2021}}; this paper, {\color{blue}{Appendix B}}) in mind, we hypothesize that if there is a causal link between the two, it is polar motion that leads atmospheric pressure. \\
    
    We have mapped the first three \textbf{iSSA} components of global sea level pressure that account for more than 85\% of the total data variance (\Figures {\color{blue}{5}},{\color{blue}{6}} and {\color{blue}{7}}). The trend had been studied in \cite{Lopes2022b}. Its pattern is remarkably stable over the 150 years of the data, and the observed geometry could be modeled as Taylor-Couette flow of mode 3 (northern hemisphere - \textbf{NH} or 4 (southern hemisphere - \textbf{SH}). The remarkable regularity and order three symmetry of the \textbf{NH} triskel occurs despite the lack of cylindrical symmetry of the northern continents. The stronger intensity and larger size of features in the \textbf{SH} is linked to the presence of the annular currents. The annual component (\Figure {\color{blue}{5}}) is characterized by a large negative anomaly extending over all of Eurasia in the \textbf{NH} Summer (and the opposite in the \textbf{NH} Winter) and three large positive anomalies over Australia and the southern tips of South America and South Africa in the \textbf{SH} Spring (and the opposite in the \textbf{SH} Autumn). The semi-annual component (\Figure {\color{blue}{6}}) is characterized by three positive anomalies (an irregular triskel) extending over Iran and the surrounding central Asia, the North-West Atlantic and the northern Pacific in the \textbf{NH} Spring and Autumn (and the opposite in the \textbf{NH} Summer and Winter), and in the \textbf{SH} Spring and Autumn by a strong stable pattern consisting of three large negative anomalies forming a clear triskel within the 40$^{\circ}$-60$^{\circ}$ annulus formed by the southern oceans (Central-South Pacific, southern Atlantic and southern Indian Ocean). A large positive anomaly centered over Antarctica, with its maximum actually displaced toward Australia, and a smaller one centered over Southern Africa complement the pattern. The pattern is opposite in the \textbf{NH} Summer and Winter. \\

    In a comparative study of variations in sea level pressure and extent of sea ice, \citep{Lopes2022c} checked that the main seasonal forcing of atmospheric oscillations was the variation of Sun-Earth distance. This is in full agreement with the changes in sign of the seasonal pressure anomalies (\Figures {\color{blue}{5}} and {\color{blue}{6}}). In Summer and Winter, annual pressure anomalies are located essentially in the northern part of the northern hemisphere, whereas in Spring and Autumn they are located on rough, elevated continental surfaces (central Asia in the northern hemisphere, the southern tips of South America, south Africa and all of Australia in the southern hemisphere). \\
    
    The present values of the ellipticity of the Earth’s orbit and the obliquity of its rotation axis lead to changes of orbital velocity (Summer in the southern hemisphere at perihelion and Summer in the northern hemisphere at aphelion). The torque changes as this velocity changes. As discussed in \citep{Lopes2021, Lopes2022a}, forces acting on the polar axis do so through the torque ($m_1$, $m_2$) obeying the Liouville-Euler system of equations. The influence of this torque is in quadrature with the length of day (\textbf{lod}). Annual variations in lod are in opposition to changes in the Sun-Earth distance ({\color{blue}{Figure 8a}}, bottom). The pressure anomalies due to the variation in rotation velocity (tracked through lod changes and winds). The phase difference between the annual components of \textbf{lod} and \textbf{SLP} is constant at 32 $\pm$ 1.2 days  (\Figure {\color{blue}{8b}}). Relative variations of amplitudes should be similar and indeed they are (\Figure {\color{blue}{8b}}). \\

 The main features that appear in the maps of \Figures {\color{blue}{5}} to {\color{blue}{7}} underline the importance of the geographical patterns of continents vs oceans and the presence of annular features at latitudes south of 20$^{\circ}$S. Three-fold symmetry is sometimes centered on or near the poles, mainly in the SH where the annular ocean ways, the southern tips of three continents and a pole-centered fourth one force this geometry. We have compared the maps of \Figures {\color{blue}{5}} to {\color{blue}{7}}  (annual, semi-annual and trend components) to the corresponding figures from \cite{Lamb1972} (Appendix C). There is in general very good agreement for the northern hemisphere. For the southern hemisphere, agreement between the two sets of maps is less. The sharp triskel patterns found in the \textbf{SSA} components are not seen as well in {\color{blue}{Lamb’s (1972)}} maps.Overall, we obtain the same paterns as {\color{blue}{Lamb’s (1972)}} from data that start around 1850 when Lamb is restricted to the 1950's, this shows the robutsness of these paterns both space and time. \\
    
    To first order, large scale atmospheric motions can be modeled as rotating cylinders \citep{Lopes2022b}, stable in time and space, in the shape of 3 or 4 mode triskels (depending on the hemispheres geography). A component of flow forced by the Sun-Earth distance modifies slightly polar rotation leading to seasonality of pressure anomalies. Since \textbf{lod} is physically linked to the solid Earth, geographical regions with strong topography are most affected by these variations. \\
    
    It is remarkable that motions are the fully explained in a frame of the theory of rotating cylinders whitout a need to call for temperature.
    
\newpage
      
\section*{Appendix A}    
\begin{figure}[H]
    \centering
	\begin{subfigure}[b]{\columnwidth}
		\centerline{\includegraphics[width=\columnwidth]{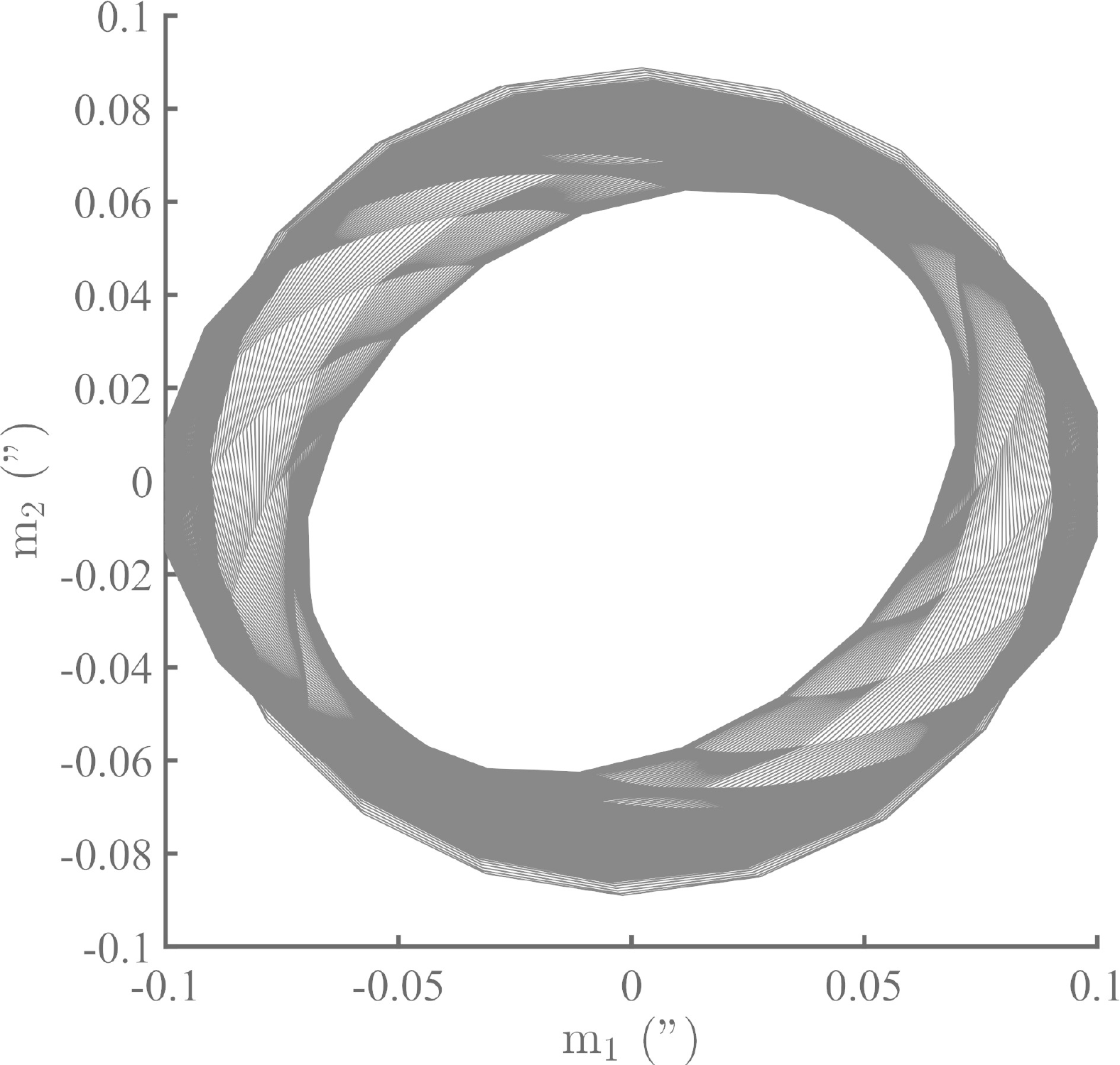}} 	
     	\subcaption{Lissajou pattern of the $m_1$ vs $m_2$ coordinates of the \textbf{iSSA} annual component of the rotation pole.}
		\label{Fig:A1} 	
	\end{subfigure}
	\begin{subfigure}[b]{\columnwidth}
		\centerline{\includegraphics[width=\columnwidth]{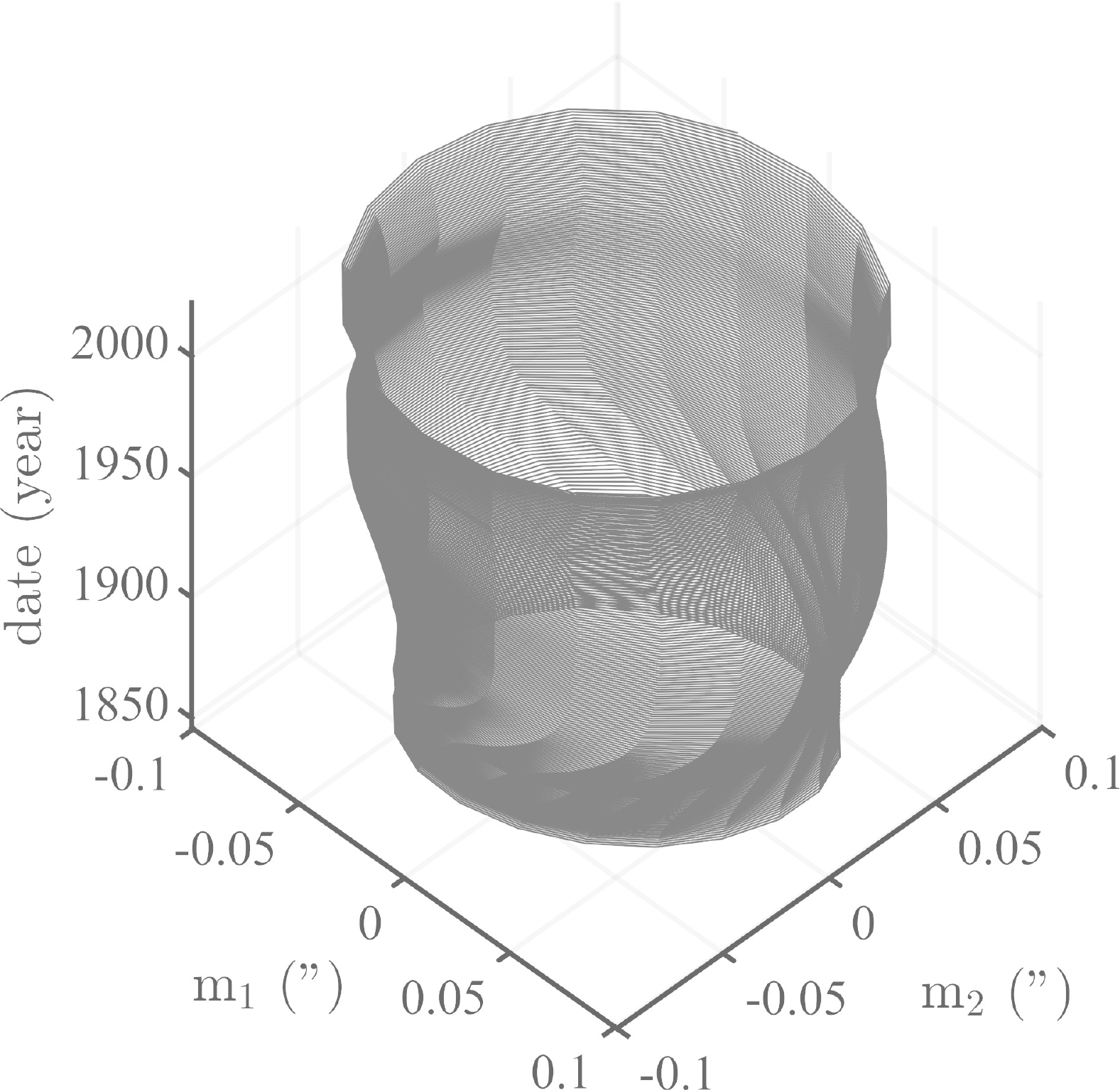}}
		\subcaption{Same as in \Figure {\color{blue}{a)}}, with the added dimension of time (1850-2020).}
		\label{Fig:A2} 	
	\end{subfigure}
	\caption*{\textit{Fig}. A: Spatio temporal evolution of the annual component of the rotation pole.}
\end{figure}	    

\newpage

\section*{Appendix B}
{\color{blue}{Laplace, 1799}}, Traité de Mécanique Céleste (volume 5, chapter 1, page 347):\\

“\textit{Nous avons fait voir (n$^{\circ}$8), que le moyen mouvement de rotation de la Terre est uniforme, dans la supposition que cette planète est entièrement solide, et l’on vient de voir que la fluidité de la mer et de l’atmosphère ne doit point altérer ce résultat. Les mouvements que la chaleur du Soleil excite dans l’atmosphère, et d’où naissent les vents alizés semblent devoir diminuer la rotation de la Terre (…) Mais le principe de conservation des aires, nous montre que l’effet total de l’atmosphère sur ce mouvement doit être insensible (…) Nous sommes donc assurés qu’en même temps que les vents analysés diminuent ce mouvement, les autres mouvements de l’atmosphère qui ont lieu au-delà des tropiques, l’accélèrent de la même quantité. On peut appliquer le même raisonnement (…) à tous ce qui peut agiter la Terre dans son intérieur et à sa surface. Le déplacement de ces parties peut seul altérer ce mouvement; si, par exemple un corps placé au pole, était transporté à l’équateur ; la somme des aires devant toujours rester la même, le mouvement de la Terre en serait un peu diminué; mais pour que cela fut sensible, il faudrait supposer de grands changement dans la constitution de la Terre}”.

\newpage
\section*{Appendix C}
\begin{figure}[H]
    \centering
	\begin{subfigure}[b]{\columnwidth}
		\centerline{\includegraphics[width=\columnwidth]{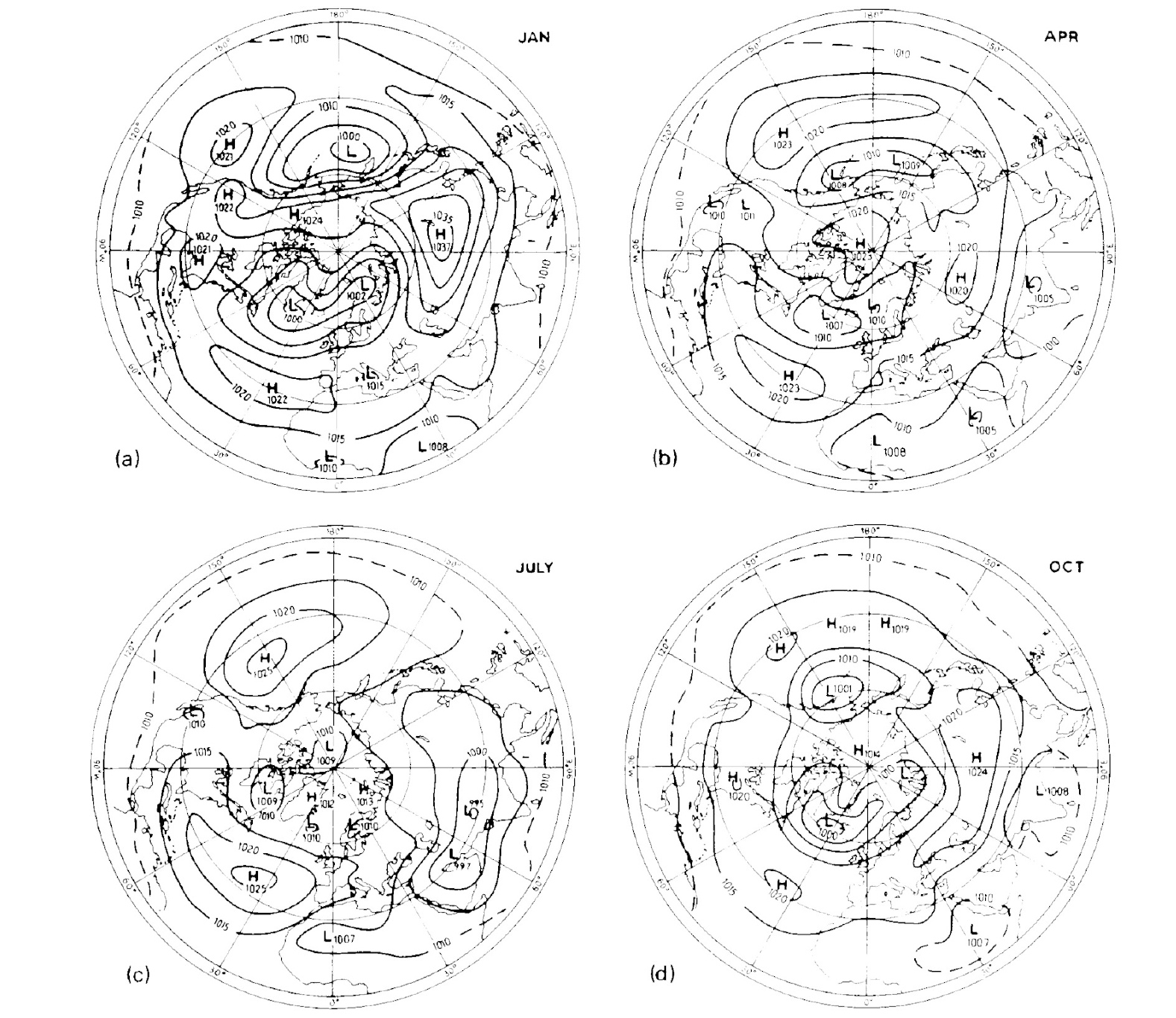}} 	
     	\subcaption{From \cite{Lamb1972} page 156, figure 4-12.  Average m.s.l. pressure over the northern hemisphere, 1950s approx: (a) January,  (b) April,  (c) July,  (d) October. Prevailing winds blow anticlockwise around the regions of low pressure, clockwise around high pressure.}
		\label{Fig:C1a} 	
	\end{subfigure}
	\begin{subfigure}[b]{\columnwidth}
		\centerline{\includegraphics[width=\columnwidth]{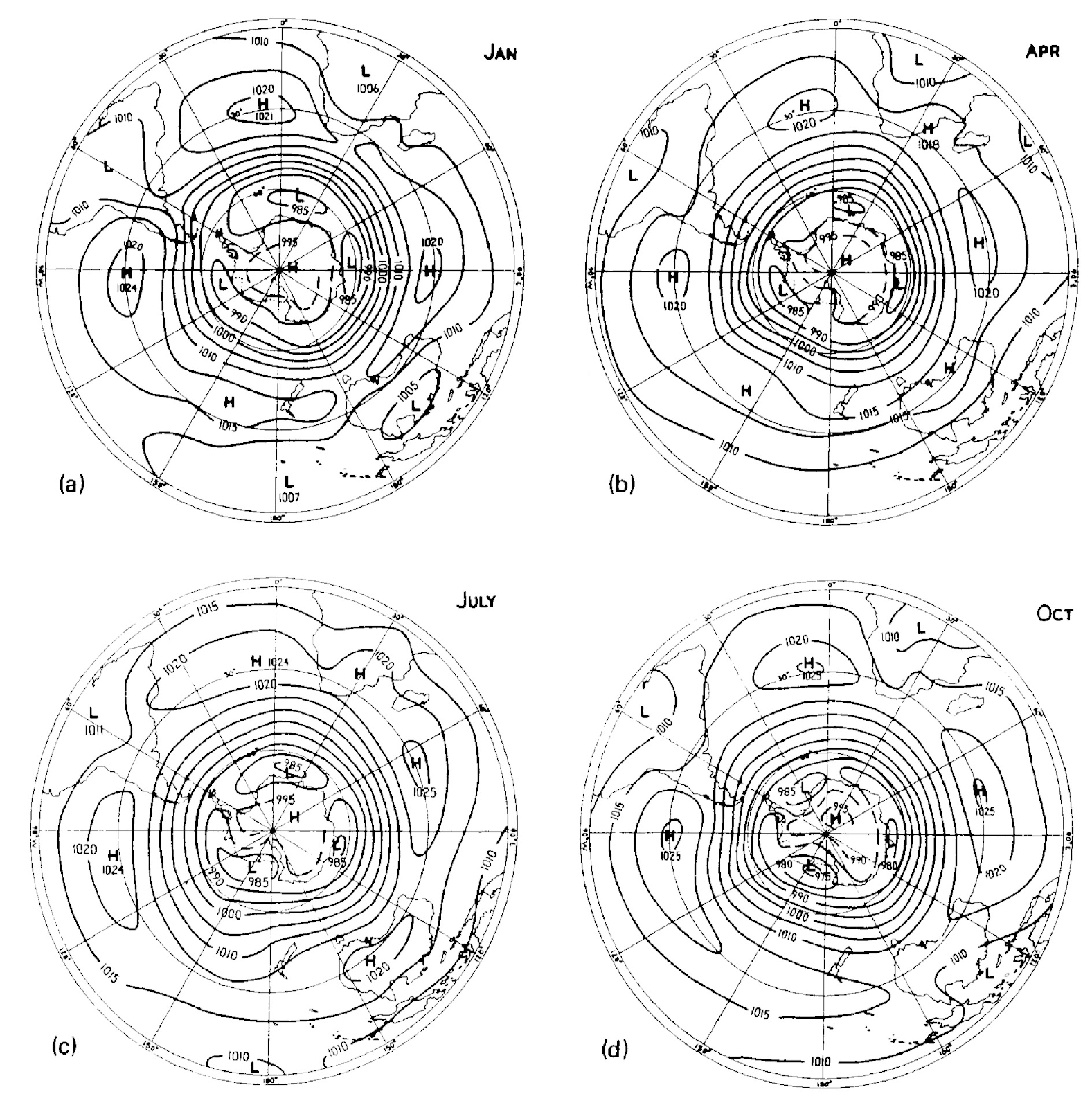}}
		\subcaption{From \cite{Lamb1972} page 157, figure 4-13.  Average m.s.l. pressure over the southern  hemisphere, 1950s approx: (a) January,  (b) April,  (c) July,  (d) October. Prevailing winds blow clockwise around the regions of low pressure, anticlockwise around high pressure.}
		\label{Fig:C1b} 	
	\end{subfigure}
	\caption*{\textit{Fig.} C1. Average m.s.l. pressure over northern (top) and the southern  hemisphere (bottom) extracted from \cite{Lamb1972}.}
\end{figure}	
\begin{figure}[H]
		\centerline{\includegraphics[width=4cm]{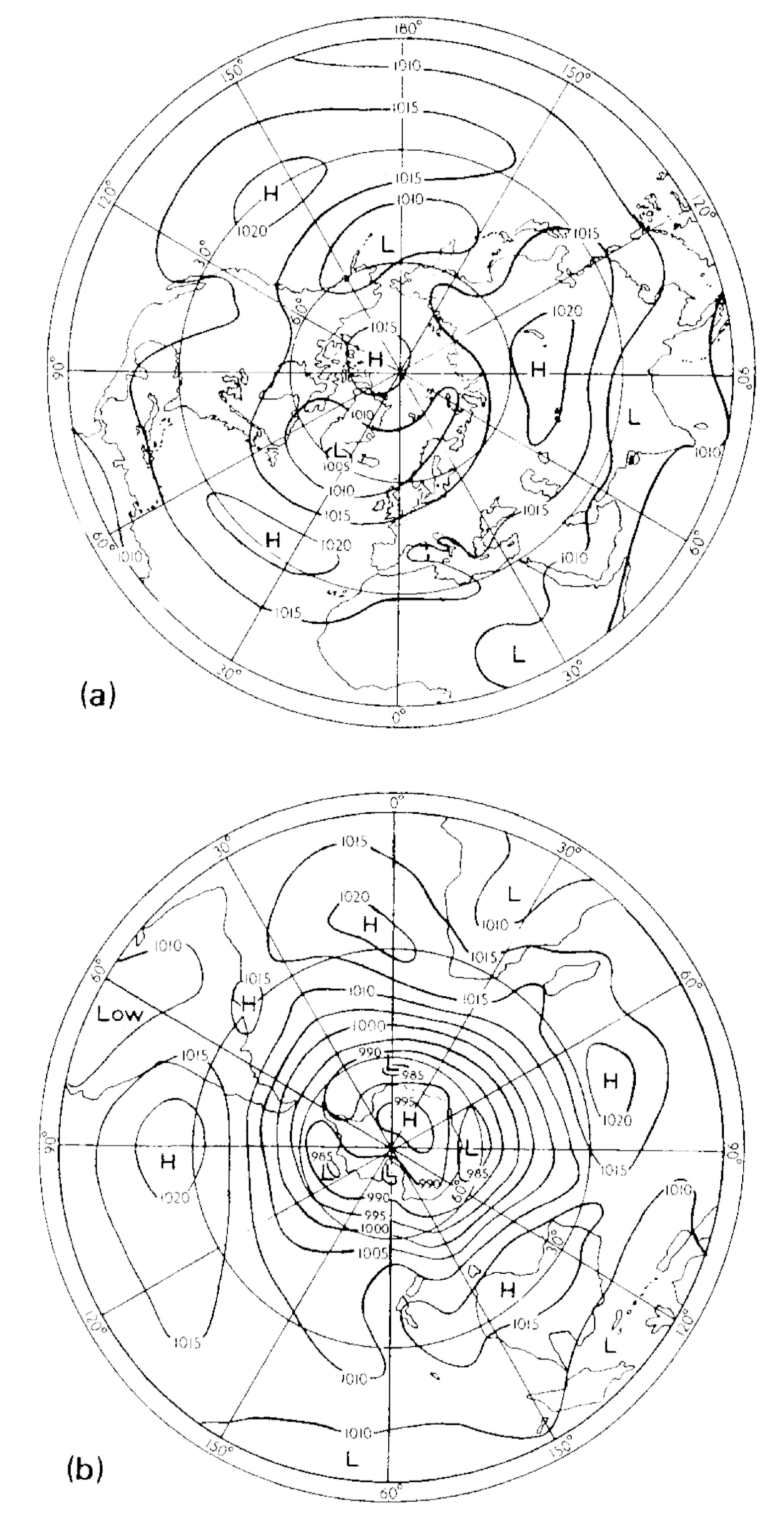}}
		\subcaption{}
		\label{Fig:C2} 	
		\caption*{\textit{Fig.} C2: From \cite{Lamb1972} page 113, figure 3-17. Annual mean distribution of atmospheric pressure in millibars at sea level: (a) northern hemisphere 1900-40 approx, (b) southern hemisphere 1900-40 approx}
\end{figure}

\bibliographystyle{aa}

\end{document}